\documentclass[a4paper,12pt]{article}
\usepackage{jheppub,esint,shuffle,psfrag}
\usepackage[utf8]{inputenc}

\usepackage{epsfig,amssymb,amsmath,psfrag,subfigure,rotate,color,wasysym,xcolor}
\usepackage[bbgreekl]{mathbbol}

\allowdisplaybreaks
\newcommand{\insertfig}[2]{\includegraphics[width=#1cm]{#2}}

\DeclareSymbolFontAlphabet{\mathbbm}{bbold}
\DeclareSymbolFontAlphabet{\mathbb}{AMSb}%

\def\XXint#1#2#3{{\setbox0=\hbox{$#1{#2#3}{\int}$ }
\vcenter{\hbox{$#2#3$ }}\kern-.6\wd0}}

\def \be  {\begin{equation}}
\def \ee  {\end{equation}}
\def \ba  {\begin{eqnarray}}
\def \ea  {\end{eqnarray}}
\def \baa {\begin{eqnarray*}}
\def \eaa {\end{eqnarray*}}
\def \lab #1 {\label{#1}}

\newcommand\re[1]{(\ref{#1})}
\def\d{\hbox{{d}\kern-.20em\hbox{l}}}

\def \qqqquad {\qquad\qquad}
\def \matrix #1 {\left(\begin{array}{cc} #1 \end{array}\right)}

\def \tr {\mathop{\rm tr}\nolimits}
\def \res{\mathop{\rm res}\nolimits}
\def \e  {\mathop{\rm e}\nolimits}
\newcommand\lr[1]{{\left({#1}\right)}}

\newcommand \vev [1] {\langle{#1}\rangle}

\def\1{\hbox{{1}\kern-.25em\hbox{l}}}

\newcommand{\ft}[2]{{\textstyle\frac{#1}{#2}}}

\makeatletter

\input pdf-trans
\newbox\qbox
\def\usecolor#1{\csname\string\color@#1\endcsname\space}
\newcommand\bordercolor[1]{\colsplit{1}{#1}}
\newcommand\fillcolor[1]{\colsplit{0}{#1}}
\newcommand\outline[1]{\leavevmode%
  \def\maltext{#1}%
  \setbox\qbox=\hbox{\maltext}%
  \boxgs{Q q 2 Tr \thickness\space w \fillcol\space \bordercol\space}{}%
  \copy\qbox%
}
\makeatother
\newcommand\colsplit[2]{\colorlet{tmpcolor}{#2}\edef\tmp{\usecolor{tmpcolor}}%
  \def\tmpB{}\expandafter\colsplithelp\tmp\relax%
  \ifnum0=#1\relax\edef\fillcol{\tmpB}\else\edef\bordercol{\tmpC}\fi}
\def\colsplithelp#1#2 #3\relax{%
  \edef\tmpB{\tmpB#1#2 }%
  \ifnum `#1>`9\relax\def\tmpC{#3}\else\colsplithelp#3\relax\fi
}
\bordercolor{black}
\fillcolor{white}
\def\thickness{.3}

\def\1{\mathbbm{1}}

\title{Octagon at finite coupling}
 
\author[a,b]{A.V.~Belitsky}
\author [b]{and G.P.~Korchemsky}
 \affiliation[a] {Department of Physics, Arizona State University, 
 Tempe, AZ 85287-1504, USA}

 \affiliation[b] {Institut de Physique Th\'eorique\footnote{Unit\'e Mixte de Recherche 3681 du CNRS}, Universit\'e Paris Saclay, CNRS, CEA, 91191 Gif-sur-Yvette} 
 
\preprint{  \parbox[t]{28mm}{IPhT--T20/007}}

 \abstract
{
We study a special class of four-point correlation functions of infinitely heavy half-BPS operators in planar $\mathcal N=4$ SYM which admit factorization into a product of two octagon form factors.
We demonstrate that these functions satisfy a system of nonlinear integro-differential equations which are powerful enough to fully determine their dependence on the 't Hooft coupling and two cross 
ratios.  At weak coupling, solution to these equations yields a known series representation of the octagon in terms of ladder integrals. At strong coupling, we develop a systematic expansion of the 
octagon in the inverse powers of the coupling constant and calculate accompanying expansion coefficients analytically. We examine the strong coupling expansion of the correlation function in various 
kinematical regions and observe a perfect agreement both with the expected asymptotic behavior dictated by the OPE and with results of numerical evaluation. We find that, surprisingly 
enough, the strong coupling expansion is Borel summable. Applying the Borel-Pad\'e summation method, we show that the strong coupling expansion correctly describes the correlation function over 
a wide region of the 't Hooft coupling. 
 }

\begin{document}

\maketitle
\flushbottom
\setcounter{footnote} 0

\section{Introduction}

In this work, we continue the study of four-point correlation functions of heavy single-trace half-BPS operators in maximally supersymmetric $\mathcal N=4$ super 
Yang-Mills theory initiated in Ref.~\cite{Belitsky:2019fan}. It has  recently been realized in Refs.~\cite{Coronado:2018ypq,Coronado:2018cxj}  that, under a
judicial choice of polarizations of  half-BPS operators, these correlation functions can be constructed at finite 't Hooft coupling $\lambda=g_{\rm YM}^2N_c$
in terms of a fundamental building block -- the octagon $\mathbb{O}$. Taking advantage of integrability of the effective two-dimensional worldsheet
 in the AdS/CFT correspondence, the octagon can be expressed in terms of the hexagon form factors \cite{Basso:2015zoa,Fleury:2016ykk,Eden:2016xvg,Bajnok:2017mdf} 
 describing dynamics of magnons propagating on the string worldsheet. 
 
 At present, explicit expressions for the octagon are known in planar $\mathcal N=4$ SYM both at weak \cite{Coronado:2018ypq,Coronado:2018cxj} and 
 strong coupling \cite{Bargheer:2019kxb,Bargheer:2019exp}.  At finite coupling, a concise representation for the octagon as a determinant of a semi-infinite 
 matrix was worked out in Refs.\ \cite{Kostov:2019stn,Kostov:2019auq}. This result laid out the foundation for our analysis in 
Ref.\ \cite{Belitsky:2019fan}, where the octagon was further recast as a Fredholm determinant of an integral operator defined on a semi-infinite line. The 
kernel of the operator in question turned out to be closely related to the Bessel kernel that previously appeared in the study of the Laguerre ensemble in 
random matrix theory \cite{Forrester:1993vtx,Tracy:1993xj}.
 
At weak coupling, the octagon is given by a multilinear combination of the so-called ladder integrals~\cite{Coronado:2018cxj}. The 
corresponding expansion coefficients can be determined to any loop order by requiring the octagon to have an appropriate asymptotic behavior in different 
kinematical limits. At strong coupling, for $g^2=\lambda/(4\pi)^2\gg 1$, the octagon possesses a semiclassical asymptotic behavior $\mathbb O\sim \e^{-g A_0+O(g^0)}$, 
where $A_0$  is the minimal area of a string that ends on four geodesics in AdS \cite{Bargheer:2019exp,Bargheer:2019kxb}. 

In this paper, we determine the octagon at finite 't Hooft coupling. To achieve this goal, we exploit its representation as a Fredholm determinant of a (modified) 
Bessel kernel. In this formulation,  the problem is akin to that encountered in study of two-point functions in integrable low-dimensional theories \cite{Korepin:1993kvr}. 
Following a general strategy after Its-Izergin-Korepin-Slavnov \cite{Its:1990}, we derive a system of exact equations obeyed by the octagon and demonstrate that its solution 
interpolates between weak and strong coupling expansions. Our consideration adds to a scarce list of examples where correlations functions in an interacting field 
theory can be determined for arbitrary coupling. 

The starting point of our analysis is a four-point correlation function of single-trace  half-BPS operators, 
dubbed the simplest in Ref.\  \cite{Coronado:2018ypq},  
\begin{align} 
\label{G4}
G_4 
= 
\vev{\mathcal{O}_1(x_1) \mathcal{O}_2(x_2) \mathcal{O}_3(x_3) \mathcal{O}_1(x_4)}
=
{ \mathcal{G}(z,\bar z) \over (x_{12}^2 x_{13}^2 x_{24}^2x_{34}^2)^{K/2}} \,,
\end{align}
where the four operators are built out of two complex scalars $Z$ and $X$ and their complex conjugate partners, $\mathcal{O}_1 =  \tr(Z^{K/2} \bar X^{K/2}) 
+ \text{permutations}$, $\mathcal{O}_2 = \tr(X^K)$ and $\mathcal{O}_3 = \tr(\bar Z^K)$. Here $\mathcal{G}(z,\bar z)$ is a function of the cross ratios
\begin{align}
\label{cross}
 {x_{12}^2 x_{34}^2\over x_{13}^2 x_{24}^2}=z \bar z\,,\qqqquad
 {x_{23}^2 x_{41}^2\over x_{13}^2 x_{24}^2}=(1-z)(1-\bar z)\,,
\end{align}
and the coupling constant $g^2 = g_{\rm\scriptscriptstyle YM}^2 N_c /(4 \pi)^2$. 

The correlation function \re{G4} is dual to a scattering amplitude of four closed string states. In the limit when the operators become infinitely heavy,  for $K\to \infty$, it factorizes 
into a double copy of an open-string partition function, the octagon \cite{Coronado:2018ypq}
\begin{align}
\label{G=O2}
\mathcal{G} (z,\bar z)  \stackrel{K \to \infty}{=} [\mathbb{O} (z,\bar z)]^2
\, .
\end{align}
Our goal is to find $\mathbb{O}$ at finite coupling constant and for generic values of the cross ratios. 

As a first step in this direction, we can consider the null limit $ x_{12}^2, x_{13}^2, x_{24}^2, x_{34}^2\to 0$, or equivalently $ z\to 0_-$ and $\bar z\to\infty$, when the 
four operators in \re{G4} are light-like separated in a sequential manner.  Introducing convenient kinematical variables $\xi = -\ft12 \log (z \bar{z})$ and $y = - \ft12 \log (z /\bar{z})$, 
one finds that in the null limit $y \gg \xi$ the octagon takes on a remarkably simple form at weak coupling \cite{Coronado:2018cxj,Kostov:2019auq}
\begin{align}
\label{WeakOcta}
\log \mathbb{O}  = -  {\Gamma(g) \over 2 \pi^2}  y^2 + {C(g) \over 8} + g^2 \xi^2 +\dots
\, ,
\end{align}
where the ellipses denote subleading terms which vanish for $y\to\infty$ with fixed $g$ and $\xi$. 

The dependence on the coupling constant enters \re{WeakOcta} through the two functions 
$\Gamma(g)$ and $C(g)$. In our previous study \cite{Belitsky:2019fan}, we used the aforementioned determinant representation of the octagon to find their exact expressions 
\begin{align} 
\label{Gam}
\Gamma(g) =   {\log (\cosh (2 \pi  g))} 
\,, \qqqquad 
C(g) =- \log \left(\frac{\sinh (4 \pi  g)}{4 \pi  g}\right)
\, .
\end{align}
Taking the relation \re{WeakOcta} at face value and going to the strong coupling limit  results in some puzzling 
consequences. As alluded to above, $\log \mathbb{O}$ is expected to have a linear growth in $g$ at most. It is easy to see that the first two terms in the right-hand side of \re{WeakOcta} do have such a
scaling whereas the third term grows as $g^2$. 

The strong coupling limit of the octagon was studied in Ref.\ \cite{Bargheer:2019exp} using a clustering technique previously developed  in Ref.\  \cite{Jiang:2016ulr} in application to three-point correlation 
functions. In the limit $g \gg y \gg \xi$, the octagon takes the form
\begin{align}
\label{InfinitGoctagon}
\log \mathbb{O} = - \frac{g}{\pi}  y^2 - g \pi + \frac{g}{\pi} \left[ \, \xi^2 \lr{\log y+1+\gamma-\log(2\pi)} + O(\xi^4)\right] + O(g^0)
\, ,
\end{align}
where $\gamma$ is the Euler-Mascheroni constant.
Comparing this relation with \re{WeakOcta} and \re{Gam}, we observe that the leading $O(y^2)$ terms coincide at strong coupling. For the second term in Eq.\ \re{WeakOcta}, however, the strong coupling limit of $C(g)/8$ 
yields a result which is twice smaller than that in \re{InfinitGoctagon}. Finally, the $\xi-$dependent terms in \re{WeakOcta} and \re{InfinitGoctagon} exhibit different dependence on $g$ and $y$. These  observations 
are troublesome, but {\it a priori\/} not completely unexpected since the two results \re{WeakOcta} and \re{InfinitGoctagon} were obtained in two different regions of the parameter space, for  $g \ll y$ and $g \gg y$, 
respectively. This suggests that the difference between \re{WeakOcta} and \re{InfinitGoctagon} at strong coupling is due to an order of limits \cite{Bargheer:2019exp}. We show below that this is indeed the case.

To address this question, we develop a  systematically expansion of $\mathbb O$ in powers of $1/g$ for general values of the cross-ratios \re{cross}. Doing so, one may attempt to apply the clustering 
technique \cite{Jiang:2016ulr}. The method is very efficient for extracting the leading large $g$ asymptotics
but it generalization to subleading terms suppressed by powers of $1/g$ proves to be difficult.~\footnote{The reason being, since the magnon rapidity scaling at strong coupling is not homogeneous, it requires 
sewing plane-wave and giant magnon regimes through the exact treatment of the near flat space domain. This is doable for a single bound state, but impenetrable for more by brute force.}  
Presently, we demonstrate that these difficulties can be avoided by solving the abovementioned system of exact equations for the octagon at large $g$.
We work out its strong coupling expansion in different kinematical limits and compare it with numerical results. We observe that the strong coupling expansion is given by a Borel 
summable series in $1/g$.  Applying the Borel--Pad\'e summation method, we show that the resulting expression for the octagon agrees with its numerical value over a wide range of the coupling constant. 
 
The paper is organized as follows. In Section~\ref{Sect2}, we review the representation of the octagon as a determinant of a semi-infinite matrix obtained in Refs.~\cite{Kostov:2019stn,Kostov:2019auq}. We 
demonstrate that this matrix can significantly be simplified by an appropriate similarity transformation allowing us to express the octagon as a Fredholm determinant of the Bessel kernel modified by some 
function of the coupling constant and cross ratios.  Applying the method of differential equations \cite{Its:1990} to this determinant, we derive a system of nonlinear integro-differential equations for the octagon. 
Its solution at weak coupling is presented in Section~\ref{sect:weak}. In Section~\ref{SectStrong}, we discuss the properties of the octagon at strong coupling and calculate the leading term of its expansion 
at large $g$. In Section~\ref{sect:strong}, we develop a systematic series of the octagon in the inverse coupling and present analytical expressions for accompanying expansion coefficients. We use these 
results in Section~\ref{sect:str} to study properties of the octagon in different kinematical regimes. In Section~\ref{SectNumerics}, we compare the obtained expressions with the numerical results. 
Section~\ref{SectConclusion} contains concluding remarks. Technical details of our analysis are summarized in four appendices. 
 
\section{Octagon as a Fredholm determinant}
\label{Sect2}

In this section, we discuss the determinant representation of the octagon and use it to derive a system of exact equations that it obeys. 

\subsection{Kinematical limits}\label{sect:lim}

We will study the octagon in Euclidean and Lorentzian kinematical regimes. Depending on the choice of the region of interest, it is convenient to introduce auxiliary kinematical variables 
\begin{align}\label{z-y}
z= \e^{-\xi+i\phi} = -\e^{-\xi-y} \,,\qqqquad \bar z= \e^{-\xi-i\phi} = - \e^{-\xi+y}
\, ,
\end{align}
where $\phi=\pi + iy$.  The variables 
 $\phi$ and $\xi$ are real in the Euclidean regime, so that $z^*=\bar z$.

The corresponding expressions for the cross ratios \re{cross} are
\begin{align}\label{zzb}\notag
& {x_{12}^2 x_{34}^2\over x_{13}^2 x_{24}^2}
= 
\e^{-2\xi }
\,,\qqqquad 
\\
&
 {x_{23}^2 x_{41}^2\over x_{13}^2 x_{24}^2}=
2  ( \cosh \xi-\cos \phi)  \e^{-\xi}
=
2  ( \cosh \xi+ \cosh y )  \e^{-\xi}
\, .
\end{align}
For future reference, we describe different OPE limits that we examine below in detail.

\subsubsection*{Euclidean  short-distance limit $x_i\to x_{i+1}$}

In this limit, two operators in the correlation function \re{G4} collide and the asymptotics of $\mathbb O$ is controlled by the operators with the lowest scaling dimensions propagating 
in the corresponding OPE channel. We can distinguish two different OPE limits
\begin{align} \label{OPE-lim}
\lr{\xi\to\infty\,,\quad \phi= 0}
\,,\qqqquad
\lr{\xi=0\,,\quad \phi\to 0}
\end{align}
corresponding to $x_1\to x_2$ and $x_1\to x_4$, respectively. 

In the first case, the leading contribution to the octagon  comes from single-trace operators with 
the scaling dimension $\Delta=K +\gamma(g)$ and it takes the form $\mathbb O\sim \e^{-\xi(\Delta-K)/2}=\e^{-\xi \gamma(g)/2}$. At strong coupling, the anomalous dimension $\gamma(g)$ 
increases with $g$ and the octagon is expected to vanish $\mathbb O\to 0$.  

In the second case, the OPE is dominated by double-trace operators with the scaling dimension $\Delta=2K$. 
Their contribution to the octagon scales as $\sim \phi^{\Delta/2-K}=\phi^0$ and the octagon approaches a finite value $\mathbb O \to 1$.

In what follows we refer to the two limits in \re{OPE-lim} as single- and double-trace  OPE limits, respectively.

\subsubsection*{Lorentzian null limit $x_{12}^2, x_{13}^2, x_{24}^2, x_{34}^2\to 0$}

In this limit, the operators in \re{G4} are located at the vertices of a null rectangle and we have 
\begin{align}
y\to\infty \,,\qqqquad \xi=\text{fixed}\,.
\end{align}
The leading contribution in each OPE channel $x_{i,i+1}^2\to 0$ comes from single-trace 
operators with the leading twist $K$ and large Lorentz spin. As in the previous case, their anomalous dimension grows indefinitely at strong coupling leading to
exponentially small octagon $\mathbb O\to 0$, see Eqs.~\re{WeakOcta} and \re{InfinitGoctagon}.

\subsubsection*{Symmetric point $y=\xi=0$}

In this case, we have $z=\bar z^*$ and $z\to -1$ in the Euclidean regime. This kinematical configuration does not correspond to a particular OPE limit. The reason why we 
consider it is that the octagon simplifies and reveals some interesting properties.
 
\subsection{Determinant representation of the octagon}

As was mentioned in the Introduction, the octagon at finite coupling admits a representation in terms of a determinant of a semi-infinite matrix \cite{Kostov:2019stn,Kostov:2019auq}
\begin{align}\label{G-ini}
\mathbb O  = \sqrt{\det\lr{1-\lambda C K}}\,,
\end{align}
where $\lambda=-2 (\cosh y+\cosh\xi)$ is a scalar factor depending on kinematical variables and the matrices $C$ and $K$ are
\begin{align}\notag\label{CK-def}
& C_{nm} = \delta_{n+1,m} - \delta_{n,m+1}
\\
& K_{mn} = -{g\over 2i}\int_{\xi}^\infty dt {\lr{i\sqrt{t+\xi\over t-\xi}}^{m-n} - \lr{i\sqrt{t+\xi\over t-\xi}}^{n-m} \over \cosh y+ \cosh t} J_m(2g\sqrt{t^2-\xi^2})J_n(2g\sqrt{t^2-\xi^2})\,,
\end{align}
where $m,n\ge 0$ and $J_m$ is a Bessel function. In the  hexagonalization approach \cite{Basso:2015zoa,Fleury:2016ykk,Eden:2016xvg},  the relation \re{G-ini} comes about  as 
a result of resummation over an arbitrary number of elementary excitations and their bound states propagating on the worldsheet of an open string describing the octagon.
 
At weak coupling, the representation \re{G-ini} can be used efficiently to expand $\mathbb O$ in powers of $g^2$. Since the matrix elements \re{CK-def} scale as $K_{nm}=O(g^{n+m+1})$, 
to any given order in $g^2$  we can replace the $K-$matrix in \re{G-ini} by its finite-dimensional minor and expand $\mathbb O$ in powers of $C K$. As was shown in Ref.~\cite{Kostov:2019auq}, 
this leads to the weak-coupling expansion of the octagon in terms of known ladder integrals as was previously bootstrapped in Refs.~\cite{Coronado:2018ypq,Coronado:2018cxj}.

At finite coupling, the calculation of \re{G-ini} may seem to be problematic due to a rather complicated form of the $K-$matrix. This matrix depends in a nontrivial way on the coupling constant 
$g$ and the kinematical variables $y$ and $\xi$ encoding the cross ratios \re{zzb}. As we demonstrated in Ref.~\cite{Belitsky:2019fan}, the octagon simplifies significantly for $\xi=0$. In this case, 
the matrix \re{CK-def} reads
 \begin{align}  \label{K0}
 (K_0)_{mn} &= -g\sin \lr{{\pi\over 2}(m-n)}
\int_{0}^\infty {dz} { J_m(2gz)J_n(2gz) \over \cosh y+ \cosh  z } \,,
\end{align}
where we inserted the subscript to indicate that this relation holds for $\xi=0$. Notice that in distinction to \re{CK-def}, the matrix elements $(K_0)_{mn}$ vanish for even $m-n$ and, as a 
consequence, the matrix $\lambda C K_0$ has a block structure. Taking advantage of this property, we were able to recast the octagon \re{G-ini} at $\xi=0$ into the form of a Fredholm 
determinant of a (modified) Bessel kernel  \cite{Forrester:1993vtx,Tracy:1993xj}. We will see momentarily that the octagon admits a similar representation for arbitrary $\xi$. 

\subsection{Similarity transformation}

Denoting $H=\lambda C K$, we observe that the octagon $\mathbb O= \sqrt{\det\lr{1-H}}$ is invariant under a similarity transformation
$H \to \Omega^{-1} H\, \Omega$. Choosing $\Omega$ appropriately we can simplify the form of the semi-infinite matrix $H$.~\footnote{The existence of such transformation was hinted in Ref.~\cite{Kostov:2019auq}.} 

As a hint, we examine a trace of this matrix, $\tr H = \lambda \tr(C K)= -2  \sum_{m\ge 0}  K_{m,m+1}$. Replacing $K_{m,m+1}$ with its expression \re{CK-def} we get
\begin{align}  \label{tr-H}
\tr H  
&= 4\lambda g^2\int_{0}^\infty {dz\,z  \over \cosh y+ \cosh(
\sqrt{z^2+\xi^2})}  \lr{J^2_0(2gz)+J_1^2(2gz) }\,,
\end{align}  
where we applied a summation formula for the Bessel functions and changed the integration variable to $z=\sqrt{t^2-\xi^2}$. Notice that, aside from the factor  $\lambda=-2 (\cosh y+\cosh\xi)$,  
the $\xi-$dependence of \re{tr-H} only resides in the denominator of the integrand. In particular, the dependence of the integrand in \re{tr-H} on $\xi$ can be restored from its value at $\xi=0$ by 
simply replacing $z\to \sqrt{z^2+\xi^2}$ in the denominator of \re{tr-H}.

Applying the same recipe to Eq.\ \re{K0}, we define the matrix
\begin{align} \label{K-Omega} 
(K_\Omega)_{mn} = -g\sin \lr{{\pi\over 2}(m-n)}
\int_{0}^\infty {dz} { J_m(2gz)J_n(2gz) \over \cosh y+ \cosh (\sqrt{z^2+\xi^2})} \,.
\end{align}
By construction, 
it satisfies the relation $\tr(CK) = \tr (CK_\Omega)$. It is a straightforward but tedious exercise 
to verify that analogous relations holds for powers of the two matrices, $\tr[(CK)^n] = \tr [(CK_\Omega)^n]$ with $n=2,3,\dots$.
This suggests that the two matrices $H=\lambda CK$ and $H_\Omega=\lambda C K_\Omega$ are related to each other by a similarity transformation
 \begin{align}\label{H-Omega}
H_\Omega= \Omega^{-1} H\, \Omega = \lambda C K_\Omega\,.
\end{align}
We verified that $\det(1-H)=\det(1-H_{\Omega})$ at weak coupling~\footnote{Analogous calculation was also performed by Valentina Petkova, we are grateful to her for sharing with us her notes.} and checked 
this relation numerically by truncating the semi-infinite matrices to a finite size and evaluating the determinants for various values of the coupling and kinematical parameters. Although we do not need the 
matrix $\Omega$ for our purposes, it would be interesting to construct it explicitly.

It is remarkable that the $\xi-$dependence of \re{K-Omega} is much simpler as compared to that of \re{CK-def}. At the same time, the matrix 
\re{K-Omega} has many properties in common with the matrix \re{K0} evaluated at $\xi=0$. In particular, the matrix elements $(H_\Omega)_{mn}$ vanish if indices $m$ and $n$ have different parity. 
The nonzero entries of this matrix with even and odd indices respectively define two  irreducible blocks. They are related to each other by a similarity transformation and, as 
a consequence, the octagon for $\xi\neq 0$ possesses the same block-diagonal form as for $\xi=0$  case (see Ref.\ \cite{Belitsky:2019fan})
\begin{align}\label{oct}
& \mathbb O= \sqrt{\det\lr{1-H_
\Omega}}=\det(1-k_-)\,.
\end{align}
Here $k_-$ is a semi-infinite matrix which is given by \re{H-Omega} with odd indices, $(k_-)_{mn} = (H_\Omega)_{2m+1,2n+1}$.
It follows from \re{H-Omega} and \re{K-Omega} that 
it admits two equivalent representations 
\begin{align}\notag\label{oct1}
& (k_-)_{nm} = (-1)^{n+m} 2 (2n+1) \int_0^\infty {dz\over z} J_{2m+1}(2gz)J_{2n+1}(2gz) \widehat\chi(z) 
\\
& \phantom{(k_-)_{nm} } =  (-1)^{n+m}  (2n+1) \int_0^\infty {dx\over x} J_{2m+1}(\sqrt x)J_{2n+1}(\sqrt x) \chi(x) \,,
\end{align}
where the notation was introduced for 
\begin{align}\label{chi}
\widehat\chi(z) = { \cosh y+\cosh\xi \over \cosh y+ \cosh (\sqrt{z^2+\xi^2})} \,,\qqqquad \chi(x) = \widehat\chi\left(\sqrt{x}\over 2g\right) \,.
\end{align}
The function $\widehat\chi(z)$ approaches $1$ at the origin and decreases exponentially fast at large $z$. It suppresses the contribution from large $z$ to \re{oct1} and serves as an ultraviolet cut-off.

The representation \re{oct} and \re{oct1} is advantageous as compared to \re{G-ini} and \re{CK-def} because the dependence of the octagon on the coupling constant and the kinematical parameters 
is confined to the cut-off function \re{chi}. We exploit this property below to formulate a system of equations for the octagon. Moreover, the relation \re{oct} can be efficiently used to compute the 
octagon numerically for arbitrary coupling (see Section~7 below). 

\subsection{Modified Bessel kernel}

Following Ref.~\cite{Belitsky:2019fan}, we can rewrite the octagon \re{oct} as a Fredholm determinant of an integral operator.
This can be achieved by expanding Eq.\ \re{oct} in terms of the traces of powers of the matrix $k_-$
\begin{align}\label{O-iter}
\log \mathbb O 
  = - \sum_{n\ge 1} {1\over n} \tr \lr{ k_-^n}  
= - \sum_{n\ge 1} {1\over n}\int_0^\infty dx_1\,\chi(x_1) \dots \int_0^\infty  dx_n \,\chi(x_n) \, K(x_1,x_2) \dots K(x_n,x_1)\,,
\end{align}
where we replaced the matrix $k_-$ by its expression \re{oct1} and introduced a notation for~\footnote{
It also admits a compact integral representation $\displaystyle K(x_1,x_2) =\frac12\int_0^1dt\, t\, J_0(t \sqrt{x_1}) J_0(t \sqrt{x_2})$. }
\begin{align} \notag\label{K-JJ}
K(x_1,x_2) &= {1\over \sqrt{x_1x_2}} \sum_{m=0}^\infty (2m+1) J_{2m+1}(\sqrt{x_1}) J_{2m+1}(\sqrt{x_2}) 
\\[2mm] 
&= {\sqrt{x_1} J_1(\sqrt{x_1}) J_0(\sqrt{x_2}) - \sqrt{x_2} J_1(\sqrt{x_2}) J_0(\sqrt{x_1})  \over 2(x_1-x_2)} 
\, .
\end{align}
This function has previously appeared in the study of level spacing distributions in random matrices and it is known as an integrable Bessel kernel~\cite{Tracy:1993xj}.  

The relation \re{O-iter} suggests to define an integral operator whose kernel is given by the Bessel kernel  \re{K-JJ} modified by the cut-off function \re{chi}
\begin{align}\label{calK}
 \mathbb{K}_\chi f(x) = \int_0^\infty {dx'} \, K(x,x')\chi(x')f(x')\,,
\end{align}
where $f(x)$ is a test function. In this way, we obtain from \re{O-iter} an equivalent representation of the octagon
\begin{align}\label{G-detK}
 \mathbb O =  \det (1-\mathbb{K}_\chi) 
\, .
\end{align}
It is remarkably similar to the analogous relation for the octagon at $\xi=0$ derived in Ref.~\cite{Belitsky:2019fan}. 

The relation \re{G-detK} holds for arbitrary values of the coupling $g$ and generic kinematical variables $y$ and $\xi$. The dependence of $\mathbb O$ on these parameters
resides in the cut-off function $\chi(x)$. This function plays a central role in our subsequent analysis. To elucidate its meaning, we examine  
\re{chi} for large $y$ and $x$. In this limit, we find that $\chi(x)$ takes the form of the Dirac-Fermi distribution
\begin{align}\label{FB}
\chi(x) \sim {1  \over 1 + \exp \lr{\frac{\varepsilon - \mu}{T}}} \,,
\end{align}
where the temperature $T$, chemical potential $\mu$ and the energy $\varepsilon$ are related to the 't Hooft coupling and kinematical variables as
\begin{align}\label{FB1}
T = 2 g\, , \qqqquad \mu = 2 g y \, , \qqqquad \varepsilon = \sqrt{x + (2 g \xi)^2} \, .
\end{align}
Replacing $x=(2g\xi)^2\sinh^2\theta$ we note that $\varepsilon = 2g\xi \cosh\theta$ coincides with the energy of a relativistic particle with mass $m=2g\xi$ and rapidity $\theta$.

Substitution of \re{FB} into \re{G-detK} and \re{calK} yields an expression for $ \mathbb O$ that resembles a Fredholm determinant representation of  two-point correlation functions 
in integrable models at finite temperature \cite{Korepin:1993kvr}.  Together with \re{FB1} this suggests that the asymptotic behavior of the octagon at weak and strong coupling should 
be similar to that of two-point correlation functions at  low and high temperature, respectively. Indeed, a  two-point function of currents in one-dimensional Bose gas model  is known to 
decay exponentially at high temperature in the so-called pre-asymptotic region, $G_2(x) \sim \exp(-T x^2/2)$ \cite{Bogoliubov:1985sjz}. The linear temperature dependence in the 
exponent translates into the linear dependence of $\log \mathbb O$ on the coupling, see Eq.~\re{InfinitGoctagon}.
  
\subsection{Method of differential equations}

A powerful technique for studying Fredholm determinants has been developed in Ref.\ \cite{Its:1990} in application to two-point correlation functions in integrable models. In our previous paper 
\cite{Belitsky:2019fan} we extended this technique to the octagon \re{G-detK} at $\xi=0$. Due to a particular form of the kernel in \re{calK}, 
generalization to arbitrary $\xi$ is straightforward.

To start with, we introduce the so-called potential $u$,  in the terminology of  \cite{Its:1990}, 
\begin{align}\label{u-def}
u  =\vev{\phi|\chi {1\over 1-\mathbb{K}_\chi}|\phi} &= \int_0^\infty dx\, dx' \, \phi(x) \chi(x) G(x,x') \phi(x')\,.
\end{align}
where $G(x,x')$ is a kernel of the operator $1/( 1-\mathbb{K}_\chi)$ and $\phi(x)= \vev{x|\phi} =J_0(\sqrt{x})$ is the Bessel function. Following \cite{Belitsky:2019fan} we can show that it is related to 
the logarithmic derivative of the octagon
\begin{align}\label{D-u}
 u = - 2 g\partial_g \log  \mathbb O\,.
\end{align}
Being a function of $g$, $y$ and $\xi$, it satisfies differential equations
\begin{align}\label{u-sys}
 \partial_\alpha u =  \int_0^\infty dx \, Q^2(x) \partial_\alpha\chi (x)\,, \qquad
 \alpha=\{g,y,\xi \}\,,
\end{align}
where the cut-off function $\chi(x)$ is given by Eq.\ \re{chi} and an auxiliary function $Q(x)$ is defined as
\begin{align}\label{Q-def} 
 Q(x)  = \vev{x|{1\over 1-\mathbb{K}}_\chi|\phi} &= \int_0^\infty dx' \, G(x,x') \phi(x')\,.
\end{align} 
The latter, in turn, obeys a differential equation
\begin{align}\label{sys}
& (g\partial_g+ 2 x\partial_x)^2 Q(x)+(x+u-g\partial_g u) Q(x) =0\,,
\end{align}
which involves the potential $u$ and its derivative with respect to the coupling constant. This equation should be supplemented with boundary conditions at weak coupling
\begin{align}\label{Q-Born}
Q(x) =J_0(\sqrt{x}) + O(g^2)\,,\qqqquad u = O(g^2)\,.
\end{align}
They follow from the expansion of \re{u-def} and \re{Q-def} in powers of $\mathbb{K}_\chi$.

The relations \re{u-sys} and \re{sys} define a system of nonlinear equations for the potential. 
Having solved them, we can determine $u$ for arbitrary $g$, $y$ and $\xi$ and, then, apply the relation \re{D-u} to compute the octagon as 
 \begin{align}\label{D-u1}
\log  \mathbb O =- \frac12 \int_0^g {d g'\over g'} u(g',y,\xi) \,.
\end{align}
This equation can be efficiently used at weak coupling whereas at strong coupling it requires knowing the potential at finite coupling.  Another relation
for $\log  \mathbb O$ was found in Ref.~\cite{Belitsky:2019fan}
\begin{align}\label{d-const}
\partial_\alpha \log\mathbb O &= \frac12 \int_0^\infty {dz \,}  \partial_\alpha \chi(x) Q^2(x)\partial_x  \lr{ g  \partial_g+2x\partial_x} \log Q(x)  \,, 
\end{align}
where $\alpha=\{g,y,\xi\}$. It can be used to determine the octagon from the solution to \re{sys}.

We can simplify the relations \re{u-sys} by taking into account the property of the cut-off function \re{chi}
\begin{align} 
\lr{2x\partial_x+g\partial_g}\chi(x)= \lr{\partial_\xi - 8g^2 \xi \partial_x - {\sinh \xi \over \cosh y+\cosh \xi}}\chi(x)=0\,.
\end{align} 
We then deduce from  \re{u-sys} 
\begin{align}\notag
\label{eqs-x}
\partial_y u &=  \int_0^\infty dx \, Q^2(x) \partial_y\chi (x)
\, ,
\\\notag
g\partial_g u &=-2  \int_0^\infty dx \, Q^2(x) x \partial_x\chi (x)
\, ,
\\
\partial_\xi u & = 8g^2 \xi \int_0^\infty dx \, Q^2(x)   \partial_x\chi (x) + {\sinh \xi\over \cosh y+ \cosh\xi} \int_0^\infty dx \, Q^2(x)  \chi (x)
\, .
\end{align} 
Being combined together with \re{D-u} and \re{sys}, these equations allow us to determine the dependence of the octagon on
the kinematical variables $y$ and $\xi$ for any value of the coupling constant $g$.

\subsection{Moments}

To analyze the relations \re{eqs-x}, it is convenient to introduce the moments
\begin{align}
\label{Qdef}
Q_\ell = - (2g)^{-2\ell} \int_0^\infty dx  \, x^\ell \, Q^2(x) \partial_{x}  \chi (x) 
\, ,
\end{align}
where the $g-$dependent normalization factor was introduced for convenience. It follows from the second relation in \re{eqs-x} that for $\ell=1$ the moment is related to the derivative of the potential
\begin{align}\label{Q1}
Q_1={1\over 8g} \partial_g u\,.
\end{align}
Moreover, it is possible to show using \re{sys} that the moments satisfy a differential equation (see Ref.~\cite{Belitsky:2019fan})
\begin{align}\label{mom-eq0}
\left[(g\partial_g)^3+ 4 (u - g\partial_g u)  (g\partial_g) - 2 g^2\partial_g^2u \right]  Q_\ell = - 16g^2 \partial_g \lr{g \,Q_{\ell+1}}
\,.
\end{align}  
It relates the moments with sequential indices. 

In particular, for $\ell=0$ the relations \re{mom-eq0} and \re{Q1} lead to the following equation for the moment $Q_0$
\begin{align}\label{mom-eq00}
\left[(g\partial_g)^3+ 4 (u - g\partial_g u)  (g\partial_g) - 2 g^2\partial_g^2u \right]  (Q_0-1) = 0
\, .
\end{align}
Obviously, $Q_0=1$ is a special solution to \re{mom-eq00}. At weak coupling, substituting $u=O(g^2)$ in \re{mom-eq00} and expanding $Q_0-1$ in powers of $g^2$, it is possible 
to show that corrections to $Q_0$ vanish to all orders in $g^2$. This suggests that for arbitrary values of $g$, $\xi$ and $y$ the solution to \re{mom-eq00} is
\begin{align}\label{Q0=1}
Q_0 =  -\int_0^\infty dx  \,  Q^2(x) \partial_{x}  \chi (x)
=1
\, .
\end{align}
We demonstrated in Ref.~\cite{Belitsky:2019fan} that for $\xi=0$ this condition leads to the expected result for the octagon \re{WeakOcta}. We argue below the same is true for $\xi\neq 0$. We will use the relation \re{Q0=1} to develop a systematic expansion of the octagon at weak and strong coupling.
   
\section{Octagon at weak coupling}\label{sect:weak}
 
To illustrate the power of the method of differential equations described in the previous section, we develop in this section the perturbative expansion of the octagon.
 
For small coupling, the octagon can be easily found from the recurrence relation for the moments \re{mom-eq0}.
It is convenient to rewrite \re{mom-eq0} as
\begin{align} \label{mom-lr}
 (g\partial_g)^3 Q_\ell = - 16g^2 \partial_g \lr{g \,Q_{\ell+1}}
-\left[ 4 (u - g\partial_g u)  (g\partial_g) - 2 g^2\partial_g^2u \right]  Q_\ell\,.
\end{align}  
Then taking into account that $u=O(g^2)$ as $g \to 0$, we observe that the expression in the right-hand side of \re{mom-lr} is suppressed by a power of $g^2$. This suggests us to solve 
\re{mom-lr} by iterations. Namely, replacing  $Q_\ell$ and $u$ in \re{mom-lr} by their perturbative expansions
\begin{align}\notag\label{u-series}
& Q_\ell = Q_\ell^{(0)} + g^2 Q_\ell^{(1)} + g^4 Q_\ell^{(2)} + \dots\,,
\\[2mm]
& u = g^2 u^{(1)} + g^4 u^{(2)} + \dots\,,
\end{align}
we compare the coefficients in front of powers of $g^2$ on both sides of \re{mom-lr} and obtain recurrence relations 
between the coefficients $Q_\ell^{(k)}$. 

Solving them, we
can express $Q_\ell$ in terms of $Q_\ell^{(0)}$ and the expansion coefficients of the potential $u^{(k)}$  
\begin{align}\label{Q-q}
Q_\ell =
q_{\ell}+g^2 \left(\frac12 {u^{(1)} q_{\ell}}-2
   q_{\ell+1}\right)+g^4 \left[ \lr{
   \frac{3}{32}(u^{(1)})^2 +\frac{3 }{8}u^{(2)}}q_{\ell}
   -\frac{3}{4} u^{(1)} q_{\ell+1}+\frac{3 }{2}q_{\ell+2} \right] + O(g^6)\,,
\end{align}
where a notation was introduced for the moments at zero coupling $q_\ell\equiv Q_\ell^{(0)}$
\begin{align}\label{mom-zero-g}
q_\ell 
& =  - \lim_{g\to 0} \ (2g)^{-2\ell} \int_0^\infty dx  \, x^\ell  J_0(\sqrt{x}) \partial_{x}  \chi (x) 
 = - \int_0^\infty dz\, z^{2\ell} \partial_z \widehat \chi(z)  \,.
\end{align}
Here in the first relation, we used the definition \re{Qdef} and replaced $Q(x)$ with its expression \re{Q-Born} at weak coupling. In the second relation, we changed the 
integration variable to $z=\sqrt{x}/(2g)$, applied \re{chi} and replaced the Bessel function by its leading asymptotics at small $g$. Replacing $\widehat \chi(z)$ in \re{mom-zero-g} with 
its explicit expression \re{chi}, we get
\begin{align}\label{q-ell}
q_\ell  = - \int_0^\infty dz\, z^{2\ell} \partial_z \left[ { \cosh y+\cosh\xi \over \cosh y+ \cosh (\sqrt{z^2+\xi^2})} \right]\,.
\end{align}
It is easy to see that $q_0=1$. For $\ell\ge 1$, we can show making use of the integration by parts that $q_\ell$ is a positive definite function of 
$\xi$ and $y$. The same is true in the Euclidean regime, for $y=i(\pi-\phi)$ and $\phi$ being real.

According to Eq.\ \re{Q1}, the moment $Q_1$ is related to the derivative of the potential with respect to the coupling. Substituting $\ell=1$ in \re{Q-q} and 
matching the result to a weak coupling expansion of $ \partial_g u/(8g)$, we can compute  the 
coefficients $u^{(k)}$.~\footnote{The same result can be obtained by requiring weak coupling corrections to vanish to $Q_0$ to all orders in $g^2$, see Eq.~\re{Q0=1}.}
The resulting expression for the potential \re{u-series} is 
\begin{align} \label{u-weak}\notag
u  {} & = 4 g^2q_1
+g^4 \left(4q_1^2-4q_2\right)
+g^6 \left(4
  q_1^3-6q_2q_1+2q_3\right)
  \\[2mm] 
  {} & + g^8 \left(4
  q_1^4-8q_2q_1^2+\frac{20}{9}q_3q_1+\frac{7
 }{3} q_2^2-\frac{5q_4}{9}\right) + O(g^{10})\,.
\end{align} 
Here the coefficients in front of the powers of $g^2$ are given by multi-linear combinations of functions \re{q-ell}. 

Note that all terms in \re{u-weak} except the first one would vanish if the functions $q_\ell$ 
had the form $q_\ell =z_0^\ell$, or equivalently the function $\widehat \chi(z)$ in \re{mom-zero-g} was given by the step function $\widehat \chi(z) = \theta(z_{0}-z)$  with some $z_0$.  Recalling \re{FB}, 
we observe that the cut-off function takes such a form in the limit of zero temperature provided that $z_0^2=y^2-\xi^2$. In this case, vanishing of the coefficients in \re{u-weak} is in agreement with the known property of the Fredholm 
determinant of the Bessel kernel at zero temperature \cite{Edelman,Forrester:1993vtx,Tracy:1993xj}.

Substituting \re{u-weak} into \re{D-u1}, we  can compute the octagon at weak coupling  
\begin{align}\notag\label{weak-O}
\mathbb O & =1-g^2 q_1+\frac{g^4 q_2}{2}-\frac{g^6 q_3}{6}+g^8
   \left(-\frac{q_2^2}{48}+\frac{q_1 q_3}{36}+\frac{5
   q_4}{144}\right)
\\ &   
   +g^{10} \left(\frac{q_2
   q_3}{144}-\frac{q_1 q_4}{96}-\frac{7
   q_{5}}{1440}\right)  
   +g^{12} \left(-\frac{q_3^2}{720}+\frac{q_2
   q_4}{2880}+\frac{7 q_1 q_{5}}{3600}+\frac{7
   q_{6}}{14400}\right) + O(g^{12})\,.
\end{align}
This relation should be compared with the analogous expression for the perturbative octagon derived in Refs.~\cite{Coronado:2018ypq,Coronado:2018cxj}.  We find that the two expressions coincide after we take 
into account that the moments \re{q-ell} can be expressed in terms of ladder integrals  
\begin{align}
q_{\ell} = {(1-z)(1-\bar z)\over z-\bar z} \sum_{m=0}^\ell (-1)^m { (2\ell-m)! \,\ell! \over  (\ell-m)!\, m! } \log^m(z\bar z) \lr{{\rm Li}_{2\ell-m}(z)-{\rm Li}_{2\ell-m}(\bar z)}\,,
\end{align}
where $z$, $\bar z$ are related to $y$, $\xi$ through \re{z-y}.

In the rest of the section, we examine \re{weak-O} in the kinematical limits described in Section~\ref{sect:lim}.

\subsubsection*{Symmetric point $\boldsymbol {y= \xi=0}$}

We find from \re{q-ell} that $q_0=1$ and $q_1=8\log 2$. For $\ell\ge 2$, the function \re{q-ell} looks as
\begin{align}\label{q-weak00}
q_\ell =   
 4 (2\ell)!  (1-2^{2-2\ell})\zeta(2\ell-1)=4(2\ell)! \, \eta(2\ell-1)\,,
\end{align}
where $\zeta$ is the Euler-Riemann zeta function and $\eta$ is the Dirichlet eta function. 
 
 The octagon \re{weak-O} admits the form
\begin{align}\notag
\mathbb O & =1-8 g^2 \log (2)+36 g^4 \zeta (3)-450 g^6 \zeta (5)
 \\[2mm]  & \notag
+g^8 \left(\frac{11025  }{2}\zeta (7)+600 \zeta (5) \log (2)-108 \zeta
   (3)^2\right)
 \\  & 
   -g^{10} \left( \frac{562275}{8} \zeta (9) +13230 \zeta (7) \log (2)-1350 \zeta
   (3) \zeta (5)\right) + O(g^{12})\,.
\end{align}
 It is convenient to assign to $\log 2$ and $\zeta(k)$ the weight $1$ and $k$, respectively. Then, 
 the first few terms of the expansion have a homogenous weight but starting from order $O(g^8)$, the coefficients have an admixture of lower-weight contributions. 

\subsubsection*{Single-trace OPE channel $\boldsymbol \xi\to\infty$ and $\boldsymbol {\phi= 0}$} 

In this case, we substitute $y=i\pi$ in \re{q-ell} and simplify the integrand at large $\xi$ to get
\begin{align}
q_\ell =  \int_0^\infty {dz\, z^{2\ell+1} \over \sqrt{z^2+\xi^2}} \e^{\xi- \sqrt{z^2+\xi^2}} = \sqrt{\frac{2}{\pi }} \, \e^{\xi } \, \xi ^{\ell+\frac{1}{2}} K_{\ell+\frac{1}{2}}(\xi )\, (2\ell) !! \,,
\end{align}
where $K_{\ell+\frac{1}{2}}(\xi )$ is the modified Bessel function. A close examination exhibits that $q_\ell$ is a polynomial in $\xi$ of degree $\ell$ with integer positive coefficients.

The corresponding expression for the octagon then reads
\begin{align}\notag\label{weak-eff}
\mathbb O &=1-2 g^2 (\xi +1)+4 g^4 \left(\xi ^2+3 \xi +3\right)-8 g^6 \left(\xi ^3+6 \xi ^2+15 \xi
   +15\right)
\\ &   \notag
   +4g^8 \left(\frac{11 \xi ^4}{3}+36 \xi ^3+159 \xi ^2+364 \xi +357\right)
\\ &  
   -8
   g^{10} \left(3 \xi ^5+43 \xi ^4+288 \xi ^3+1104 \xi ^2+2385 \xi
   +2295\right) + O(g^{12})\,.
\end{align}
Note that the expansion goes in powers of $g^2\xi$ and $1/\xi$.

\subsubsection*{Double-trace OPE channel $\boldsymbol \xi=0$ and $\boldsymbol {\phi\to  0}$} 

Replacing $y=i(\pi-\phi)$ in \re{q-ell} and going through the calculation of the integral at small $\phi$, we get $q_0=1$  and 
\begin{align}\notag\label{q1}
& q_1=2\phi^2 (1-\log\phi) + O(\phi^4)\,,
\\[2mm]
& q_\ell=(2\ell)! \zeta(2\ell-1) \phi^2 + O(\phi^4)\,,\qquad \text{for $\ell\ge 2$}\,.
\end{align}
In this case, the octagon is given by
\begin{align}\label{log-phi}
 \mathbb O
=1+ \phi^2 \left[ 2 g^2 (\log \phi -1)+12 g^4 \zeta (3)-120 g^6 \zeta (5) 
+ O(g^{8})\right] + O(\phi^4).
\end{align}
The expansion coefficients of $\mathbb O$ depend on $\log\phi$. Such terms come from $q_1$ in \re{q1}. Since the expression on the right-hand side of 
 \re{weak-O} is linear in $q_1$, the expansion coefficients are linear in $\log\phi$ to any order in $g^2$~\cite{Coronado:2018cxj}.  

Notice that the weak-coupling corrections to $\mathbb O$ vanish as $\phi\to 0$. In this limit, the leading contribution to the octagon comes from double-trace half-BPS operators 
with the scaling dimension $K$ propagating in the OPE channel $x_{14}^2\to 0$. It is protected from quantum effects and leads to $ \mathbb O\to 1$ as $\phi\to 0$.
 
\subsubsection*{Null limit $\boldsymbol \xi=\text{fixed}$ and $\boldsymbol {y\to \infty}$} 
 
 At large $y$, the dominant contribution to \re{q-ell} arises from $z=O(y)$. Replacing the integration variable $z\to y z$, we get from \re{q-ell} in the large-$y$ limit
\begin{align}\notag
q_\ell
& =  y^{2\ell} \int_0^\infty dz\, z^{2\ell} (-\partial_z){1\over 1+ \e^{\sqrt{y^2z^2+\xi^2}-y}} 
\\
& = (-\xi^2)^{\ell} + 2\ell\sum_{k=0}^{\ell-1}(-\xi^2)^{\ell-1-k}  \lr{\ell-1\atop k} B_{2k+2}\lr{\frac12 +{iy\over 2\pi}} {(-2i \pi)^{2k+2}\over 2k+2}\,,
\end{align}
where $B_{2k+2}(x)$ is a Bernoulli polynomial. The function $q_\ell$ is a polynomial in $y^2$ of degree $\ell$.  

In this case, it is advantageous to consider the logarithm of the octagon \re{weak-O}
\begin{align}\notag\label{nullO}
\log \mathbb O &= -g^2 \left(y^2-\xi ^2+\frac{\pi ^2}{3}\right)+g^4 \left(\frac{2 \pi ^2 y^2}{3}+\frac{8 \pi
   ^4}{45}\right)-g^6 \left(\frac{32}{45} \pi ^4 y^2+\frac{512 \pi ^6}{2835}\right)
   \\ & 
   +g^8
   \left(\frac{272 \pi ^6 y^2}{315}+\frac{1024 \pi ^8}{4725}\right)-g^{10}
   \left(\frac{15872 \pi ^8 y^2}{14175}+\frac{131072 \pi ^{10}}{467775}\right) + O(g^{12})\,.
   \end{align}
This relation is in agreement with \re{WeakOcta} and \re{Gam}.  In particular, the expansion coefficients of $\log \mathbb O$ are linear in $y^2$ and the $\xi-$dependence 
cancels to all orders except the lowest one. 

The last property can be understood using the last relation in \re{eqs-x}. The first term in its right-hand side is proportional to the moment $Q_0$. Taking into account \re{Q0=1}, we get
\begin{align}\label{neg}
\partial_\xi u & = - 8 g^2 \xi + {\sinh \xi\over \cosh y+ \cosh\xi} \int_0^\infty dx\, Q^2(x)  \chi (x) \,.
\end{align}
We can show that the second term in this equation is exponentially small at large $y$. Indeed, at small $g$, or equivalently at low temperature, the function $\chi(x)$ effectively reduces the 
integration region in \re{neg} to $x< (2gy)^2$. Replacing $Q(x)=J_0(\sqrt{x})+O(g^2)$ we find that the integral in \re{neg} scales as $ 4g^2 y^2 + O(g^4)$. It is accompanied, however, by the 
factor of  ${\sinh \xi/ (\cosh y+ \cosh\xi)}\sim \e^{-y}$ and, therefore, provides a vanishing contribution to \re{neg} at large $y$. Thus, $\partial_\xi u = - 8 g^2 \xi$ at weak coupling leading to 
$\partial_\xi \log  \mathbb O = 2 g^2 \xi  $ in virtue of \re{D-u1}.

As was mentioned in Section~\ref{sect:lim}, the relation \re{nullO} describes the asymptotic behavior of the null octagon. It arises as a result of infinite resummation of contributions of leading twist 
operators in different OPE channels. Exponentially small corrections, generated by the second term in the right-hand side of \re{neg} at weak coupling, are induced by high twist operators. 
We show below that at strong coupling the situation is completely different. Anomalous dimensions of the operators grow with the coupling and their classification with respect to twist ($=$ bare 
dimension minus spin) becomes redundant. We demonstrate in Section~\ref{sect:str} that the second term in \re{neg} scales at strong coupling as $8g^2\xi +O(g)$, so that $u=O(g)$ in 
agreement with the expected behavior of the octagon \re{InfinitGoctagon}.
 
\section{Octagon at strong coupling}
\label{SectStrong}

In this section, we study the octagon in the limit when $g\to\infty$ with the kinematical variables $y$ and $\xi$ held fixed.

In this limit,  the octagon is expected to have the following form
\begin{align}\label{O-str}
\log \mathbb O = - g A_0 + \frac12 A_1^2\, \log g + B + {A_2\over 4g} + {A_3\over 12 g^2}+ {A_4\over 24 g^3} + \dots\,,
\end{align}
where the coefficient functions depend on the kinematical variables $y$ and $\xi$ and relative rational prefactors are introduced for convenience. The sum contains terms of the form $A_k g^{1-k}/(2k(k-1))$.

The first term in \re{O-str} was computed in Ref.~\cite{Bargheer:2019exp} using the clustering procedure developed in Ref.~\cite{Jiang:2016ulr} (see Eq.~\re{A0-ref} below). 
It was argued there that  $A_{0}$ should be related to the minimal area of a string that ends on four geodesics in AdS and rotates on the sphere. 
The remaining terms in  \re{O-str}  remain unknown.

The subleading term $A_1^2\, \log g/2 + B$ in \re{O-str} describes quadratic fluctuations around the minimal area. It is enhanced by $\log g$ and produces an overall $g-$dependent factor of the 
form~\footnote{Defining $\log g$ term in \re{O-str}, we anticipated the power of $g$ to be positive in this relation.}
\begin{align}
\mathbb O =g^{A_1^2/2} \e^{-g A_{0} + B+ O(1/g)}\,.
\end{align}
Here $O(1/g)$ terms in the exponent take into account yet higher order quantum fluctuations. 

We are going  
to determine the coefficient functions in \re{O-str} from the system of equations \re{D-u}, \re{eqs-x} and \re{sys}. Combining together \re{O-str} and \re{D-u} we expect the potential 
to have the following form at strong coupling
\begin{align}\label{u-str}
u= - 2 g\partial_g \mathbb O =  2 gA_{0} - A_1^2 + {A_2\over 2g} + {A_3\over 3g^2} + {A_4\over 4 g^2} + \dots\,.
\end{align}
Comparing this relation with \re{O-str} we notice that $B$ does not contribute to $u$. This means that having determined the potential $u$, we will be able to determine the octagon \re{O-str} up to the function $B(y,\xi)$. 

\subsection{Leading order}

To find the leading term $A_0$, we use the relation \re{O-iter} and examine the properties of its iterated integral representation in the limit $g\to\infty$. 

It follows from the explicit form of the cut-off function \re{chi} that the dominant contribution to \re{O-iter} comes from integration over $x_i=O(g)$. This allows us to replace the $K-$kernel in 
\re{O-iter}  by its leading asymptotic behavior at large $x_i$
\begin{align}\label{K-exp}
K(x_1,x_2) ={1\over 2 \pi   (x_1 x_2)^{1/4}}  \left[\frac{ \sin
   \left(\sqrt{x_1}-\sqrt{x_2}\right)}{ 
    \sqrt{x_1}-\sqrt{x_2}}-\frac{ \cos \left(\sqrt{x_1}+\sqrt{x_2}\right)}{  \sqrt{x_1}+\sqrt{x_2} }\right] + \dots\,.
\end{align}
Here the ellipses denote subleading terms suppressed by powers of $1/x_i$. We notice that for $x_i=O(g)$, the second term inside the square brackets is a rapidly oscillating function of $x_i$. At large $g$ it scales as 
$O(1/g^{1/2})$. At the same time, the first term is peaked around $x_1=x_2$ and scales as $O(g^0)$ for $x_1-x_2=O(1/g)$. This suggests that for $g\to\infty$, the integral in \re{O-iter} is localized 
at $x_1=x_2=\dots=x_n$. 

To show this, we use the identity
\begin{align}\notag\label{K-int}
\int_0^\infty dx_2 \,K(x_1,x_2) \chi(x_2) f(x_2) 
& = \int_0^\infty dx_2 \frac{\sin\left(\sqrt{x_1}-\sqrt{x_2}\right)}{ 2 \pi   (x_1 x_2)^{1/4}(\sqrt{x_1}-\sqrt{x_2})} \chi(x_2) f(x_2) +\dots
\\
& =    \int_{-\infty}^\infty dz \frac{\sin(z)}{\pi z} \chi(x_1) f(x_1) +\dots = \chi(x_1) f(x_1) +\dots\,,
\end{align}
where $f(x)$ is a smooth test function. Here in the second relation, we changed the integration variable to $z=\sqrt{x_2} -\sqrt{x_1}$ and expanded the integral at large $x_1$ and fixed $z$. 

Subsequently applying \re{K-int}, we get from  \re{O-iter}  
\begin{align}  \label{O-iter1}
\log \mathbb O & = - \sum_{n\ge 1} {1\over n}\int_0^\infty dx_1\,[\chi(x_1)]^n  \, K(x_1,x_1)  = \int_0^\infty dx_1\,\log(1-\chi(x_1)) K(x_1,x_1)\,.
\end{align}
As before, we can replace $K(x_1,x_1)$ with its leading asymptotic behavior at large $x_1$. Applying \re{K-exp} for $x_2\to x_1$, we get
$K(x_1,x_1)\sim 1/(2\pi\sqrt{x_1})$. Then, changing the integration variable to $x_1=(2g z)^2$, we finally obtain in the leading large $g$ limit
\begin{align} \label{O-LO}
\log \mathbb O & ={2g\over\pi}\int_0^\infty dz \log(1-\widehat\chi(z)) + O(g^0)\,.
\end{align}
According to its definition \re{chi}, the function $\widehat\chi(z)$ is independent of the coupling constant but carries  dependence on the kinematical variables $y$ and $\xi$.

Comparing \re{O-LO} with \re{O-str}, we deduce that
\begin{align}\label{A-1}
A_0 
& = -{2\over\pi}\int_0^\infty dz \log(1-\widehat\chi(z))
 =  {1\over\pi} \int_{-\infty}^\infty dz \log\lr {\cosh (\sqrt{z^2+\xi^2}) + \cosh y \over \cosh (\sqrt{z^2+\xi^2})-\cosh\xi}\,.
\end{align}
As follows from this relation, $A_0$ is a positive definite function of real $y$ and $\xi$. The same is true in the Euclidean regime for $y=i(\pi-\phi)$ with $\phi$ real.

As was mentioned at the beginning of this section, the function $A_0$ was also computed in Ref.~\cite{Bargheer:2019exp} using a different technique. To compare the two results, we change the 
integration variable in \re{A-1} to $z=\xi\sinh \theta$ and introduce a new function  
\begin{align} 
 Y(\theta) & = -\widehat\chi(z) = - {\cosh \left(\ft12(\xi  + y) \right) \cosh \left(\ft12(\xi  - y) \right) \over \cosh \left(\ft12(\xi\cosh\theta + y) \right) \cosh \left(\ft12(\xi\cosh\theta - y) \right)}\,.
\end{align}
Then, the relation \re{A-1} takes the form
\begin{align}\label{A0-ref}
A_0 =  -  \int_{-\infty}^\infty  {d\theta\over \pi} \, \xi \,  \cosh\theta\, \log(1+Y(\theta))
\end{align}
and it coincides with the analogous expression obtained in Ref.~\cite{Bargheer:2019exp}.

\subsection{Beyond leading order} 
   
To derive the strong coupling expansion \re{O-str}, we use the system of integro-differential equations \re{D-u}, \re{sys} and \re{eqs-x}. Namely, we will construct the function $Q(x)$ at strong 
coupling for arbitrary $u$ and, then, apply the second relation in \re{eqs-x} to determine the expansion coefficients of the potential in \re{u-str}.

The function $Q(x)$ depends on the coupling constant and obeys Eq.\ \re{sys}.
It is convenient to replace $x=(2gz)^2$ and introduce the function 
\begin{align}\label{q=Q}
q(z) = Q(4g^2z^2)\,.
\end{align}
It follows from \re{sys} that it satisfies the following differential equation
\begin{align}\label{diffeq}
\left[ (g\partial_g)^2+ 4g^2 z^2 + w(g) \right] q(z) = 0  \,,
\end{align}
where a notation was introduced for
\begin{align}\label{w}
w(g) = u - g \partial_g u=-A_1^2 +{A_2\over g} + {A_3\over g^2} + {A_4\over g^2} + \dots
\end{align}
and  the potential $u$ was replaced with its general expression at strong coupling \re{u-str}. Note that the leading $O(g)$ term in \re{u-str} proportional to $A_0$ does not contribute to $w(g)$.
 
Substituting \re{q=Q} into \re{eqs-x}, we get the system of equations relating the potential and solutions to \re{diffeq}
\begin{align}\notag\label{eqs}
\partial_y u &=  8g^2 \int_0^\infty dz \, z\, q^2(z) \partial_y\widehat \chi (z)\,,
\\\notag
 \partial_g u &=-8g \int_0^\infty dz \, z^2\, q^2(z) \partial_z\widehat \chi (z)\,,
\\
\partial_\xi u & =  8 g^2 \xi  \int_0^\infty dz  \, q^2(z) \partial_{z}\widehat \chi (z) + {8g^2 \sinh \xi\over \cosh y+ \cosh\xi} \int_0^\infty dz \,z\, q^2(z)  \widehat \chi (z) \,,
\end{align}  
with $\widehat \chi(z)$ defined earlier in Eq.\ \re{chi}. In a similar manner, it follows from \re{Q0=1} that
\begin{align}\label{Q0-mom}
Q_0 = \int_0^\infty dz  \,  q^2(z) (-\partial_{z}\widehat \chi (z)) =1\,.
\end{align}
Solution to \re{diffeq} yields the function $q(z)$ that depends on $u(g)$ in a nontrivial way. Its substitution into \re{eqs} and \re{Q0-mom} yields a complicated system of equations for the potential.  
Luckily, these equations can be solved at strong coupling. 

\section{Strong coupling expansion}\label{sect:strong}

In this section, we use the relations \re{diffeq} -- \re{Q0-mom} to systematically calculate the coefficients $A_1$, $A_2$, $A_3,\dots$ of the strong coupling expansion of the octagon \re{O-str}.

\subsection{Next-to-leading order}
 
To begin with, we neglect the $O(1/g)$ correction to \re{w} and examine the differential equation \re{diffeq}. In this case,  for $w=-A_1^2$, a general solution to \re{diffeq} reads
\begin{align}\label{q-c}
q(z) = c (z) J_{A_1}(2gz) + c'(z) Y_{A_1}(2gz) + O(1/g)\,,
\end{align}
where $J_\alpha$ and $Y_\alpha$  (with $\alpha=A_1$) are Bessel functions of the first and second kind, respectively, and $c(z)$, $c'(z)$ are arbitrary functions of $z$. The last term in 
the right-hand side of \re{q-c} denotes corrections due to $O(1/g)$ terms in \re{w}. For arbitrary $\alpha$, the function $Y_\alpha$ has singularity at the origin and is multivalued. This 
suggests to put $c'(z)=0$. 

To find the function $c (z)$, we examine the second relation in \re{eqs}
\begin{align} \label{u-chain}
 \partial_g u &=-8g \int_0^\infty dz \, z^2\, q^2(z) \partial_z\widehat \chi (z)  
 =2 A_{0} + O(1/g^2)\,,
\end{align}
where in the last relation we replaced $u$ with its expression \re{u-str}.
Evaluating the integral in \re{u-chain}, we can replace the function $q(z)$ with its leading asymptotic behavior
at large $g$,
\begin{align}
q^2(z) =  c^2(z) [J_{A_1 }(2gz)]^2 =  c^2(z)\frac{1+\sin\big(4 g z-\pi A_1\big)}{2\pi  g z} + \dots \, .
\end{align}
Finally, substituting this expression in the previous relation, we deduce
\begin{align} \label{der-u}
 \partial_g u 
 &=-{4\over\pi} \int_0^\infty dz \, z  \partial_z\widehat \chi (z)  c^2(z)  + \dots \, .
\end{align}
Here we took into account that the dominant contribution to the integral comes from $z=O(g^0)$ and, as a consequence,
a rapidly oscillating sinus function does not contribute. Matching the last relation to the right-hand side of \re{u-chain} and replacing $A_0$ with \re{A-1}, we conclude
that $c^2(z) = {1/(1-\widehat \chi(z))}$. Thus, the solution \re{q-c} looks as
\begin{align}\label{q-c1}
q(z) = {1 \over [1-\widehat \chi(z)]^{1/2}} \left[ J_{A_1}(2gz) + O(1/g)\right]\,.
\end{align}
Here the second term in the numerator denotes corrections due to $O(1/g)$ terms in \re{w}.
Substituting \re{q-c1} into \re{eqs} and repeating the same analysis, it is straightforward to verify that the two remaining relations in \re{eqs} are automatically satisfied. 

Having constructed the function $q(z)$, we can examine now the normalization condition \re{Q0=1}
\begin{align} \label{Q0-int} 
Q_0
& =  \int_0^\infty dz\, [J_{A_1 }(2gz)]^2 {(-\partial_z \widehat \chi(z))\over 1-\widehat \chi(z)} = 1
\, .
\end{align}
In distinction to \re{der-u},  the leading contribution to this integral comes from $z=O(1/g)$. In this region, we can replace the cut-off function \re{chi} by its leading behavior  around the origin
$
 \partial_z  \log(1-\widehat \chi(z)) = 2/z + \dots
$.
We can thus continue the previous relation to get
\begin{align}
 Q_0=2\int_0^\infty {dz\over z}\, [J_{A_1 }(2 gz)]^2 ={1\over A_1} =1 \,.
\end{align}
We conclude that 
\begin{align}\label{A1}
A_1=1\,.
\end{align}
Thus, the subleading $O(\log g)$ correction to the octagon \re{O-str} is universal, i.e., it does not depend on the kinematical variables $y$ and $\xi$.

\begin{figure} 
\psfrag{f}[cr]{$f(z)\ \ $}
\psfrag{z}{$z$}
\centerline{\parbox[c]{82mm}{\insertfig{8.2}{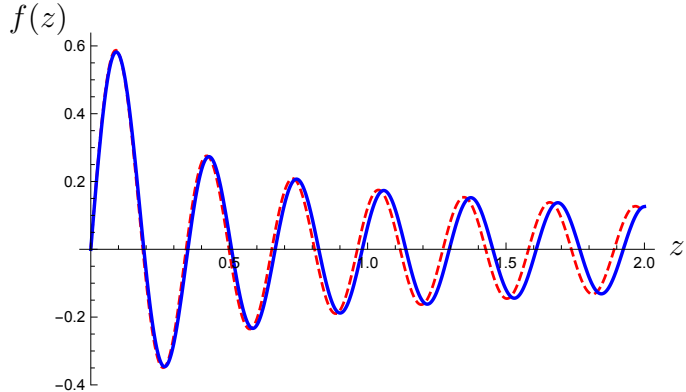}}}
\caption{The function $f(z)$ defined in \re{q-gen}. 
Blue curve was obtained from \re{q=Q} and \re{Q-rep}  at $g=10$, $y=1$ and $\xi=1/10$ by truncating an infinite sum in \re{Q-rep} to the first $N_{\rm max}=100$ terms. 
Dashed red line depicts the leading term $J_{1}(2gz)$. }
\label{fig:q-fun}
\end{figure}

Combining together \re{q-c1} and \re{A1}, we find that the solution to the differential equation \re{eqs} looks as
\begin{align} \label{q-gen}
q(z) = {1 \over [1-\widehat \chi(z)]^{1/2}} f(z)\,,\qqqquad
f(z) = J_{1}(2gz) + O(1/g)\,.
\end{align} 
We can check this relation by computing the function $q(z)$ numerically 
at some reference value of the coupling and the kinematical parameters as described in Appendix~\ref{app:mat}. 
Its comparison with the leading term in \re{q-gen} is shown in Figure~\ref{fig:q-fun}.

It is important to stress that at strong coupling the integrals in \re{u-chain} and \re{Q0-int} receive the leading contributions from two different regions, $z=O(g^0)$ and $z=O(1/g)$, respectively. 
This suggests that, in order to compute the corresponding expressions for $\partial_g u$ and $Q_0$, we have to construct the function \re{q-gen} in these two regions. For $z=O(g^0)$ this is done in the next subsection and for $z=O(1/g)$ in Appendix~\ref{app:sol}.

\subsection{Solution at finite $z$}  

At large $g$ and $z=O(g^0)$, it is convenient to switch from $g$ to $x= gz$ and expand the function \re{q-gen} in powers of $1/x$. Since the function $f$ differs from $q(z)$ by a $g-$independent factor, 
it satisfies the very same differential equation \re{diffeq} 
\begin{align}\label{PDF1}
\left[(x\partial_x)^2 + 4 x^2 -1 + A_2 { z\over x} +A_3 { z^2\over x^2} + A_4 { z^3 \over  x^3} +\dots\right] f  =0\,,
\end{align}
where we used \re{w} and replaced $g=x/z$. 

To leading order in $1/x$, the solution is given by \re{q-gen}. 
Using the properties of Bessel functions, we find from \re{q-gen} that $f$ is an oscillating function 
\begin{align}\label{hat-q-as}\notag
f  ={1\over \sqrt{2\pi x}}\Big[& \left(1+\frac{3}{16 x}+\frac{15}{512 x^2}+\dots \right) \sin (2 x)  
\\
& + \left(-1+\frac{3}{16 x}-\frac{15}{512
   x^2}+\dots \right) \cos (2 x)  \Big] + O(z/x)\,.
\end{align} 
Going beyond leading order, we look for a general solution to \re{PDF1} in the form
\begin{align}\label{hat-q-ansatz}
f = {1\over \sqrt{2\pi x}}\left[a(x,z) \sin (2 x)+ b(x,z)\cos (2 x) \right]
 ,
\end{align}
where $a$ and $b$ are given by infinite series in $1/x$
\begin{align}
a(x,z)= \sum_{n=0}^\infty  {a_n(z)\over x^n} \,,\qqqquad b(x,z)= \sum_{n=0}^\infty  {b_n(z)\over x^n} 
\, .
\end{align}   
Substituting \re{hat-q-ansatz} into \re{PDF1} and matching the coefficients in front of powers of $1/x$, we can determine the functions $a(x,z)$ and $b(x,z)$ to any order in  $1/x$
\begin{align} \label{A}\notag
a(x,z)& =1+\frac{3}{16 x}+\lr{-\frac{A_2 z}{8}+\frac{15}{512}}{1\over x^2}+\lr{-\frac{1}{12} A_3
   z^2-\frac{5 A_2 z}{128}-\frac{105}{8192}}{1\over x^3}
 \\  
  & +\lr{-\frac{1}{16} A_4
   z^3-\frac{1}{128} A_2^2 z^2-\frac{3 A_3 z^2}{64}+\frac{105 A_2
   z}{4096}-\frac{4725}{524288}}{1\over x^4}+O(1/x^5)\,,
\\ \notag \label{B}
b(x,z) & = -1+\frac{3}{16 x}+\lr{-\frac{A_2 z}{8}-\frac{15}{512}}{1\over x^2}+\lr{-\frac{1}{12} A_3
   z^2+\frac{5 A_2 z}{128}-\frac{105}{8192}}{1\over x^3}
 \\ &  
   +\lr{-\frac{1}{16} A_4
   z^3+\frac{1}{128} A_2^2 z^2+\frac{3 A_3 z^2}{64}+\frac{105 A_2
   z}{4096}+\frac{4725}{524288}}{1\over x^4}+O(1/x^5)\,.
\end{align}
As a check, we verify that for $z=0$ the functions $a(x,0)$ and $b(x,0)$ coincide with the coefficients accompanying trigonometric functions inside the brackets in Eq.\ \re{hat-q-as}.
  
Let us now examine \re{u-chain} and replace the function $q(z)$ with its expression \re{q-gen} and \re{hat-q-ansatz}  
\begin{align} \label{u-chain1}
 \partial_g u &= {4\over \pi} \int_0^\infty  dz \left[a(2gz,z) \sin (2 gz)+ b(2gz,z)\cos (2 gz) \right]^2 z \partial_z \log(1-\widehat \chi(z) ) \,.
\end{align}
We recall that the cut-off function \re{chi} does not depend on the coupling constant and the dependence on $g$ resides in the first factor in the right-hand side of \re{u-chain1}. At strong coupling 
this factor contains rapidly oscillating trigonometric functions. Expanding the integral at large $g$, we can safely replace them by their averaged values. This leads to
\begin{align}\label{u-der1}
 \partial_g u = {2\over \pi} \int_0^\infty  dz \, z \partial_z \log(1-\widehat \chi(z) ) \left[a^2(2gz,z) + b^2(2gz,z)\right] \,.
\end{align}
It is important to stress that this relation was derived under the assumption that the dominant contribution to the integral comes from $z=O(g^0)$.  

Using \re{A} and \re{B}, we find
\begin{align}\notag\label{AB}
& \frac12\left[a^2(2gz,z) + b^2(2gz,z)\right] = 1+\frac{3}{32 z^2 g^2 }-\frac{A_2}{8 z^2 g^3}-\lr{\frac{45}{2048
   z^4}+\frac{A_3}{8 z^2}}{1\over g^4}
 \\  
 & \qquad +\lr{\frac{15 A_2}{256 z^4}-\frac{A_4}{8 z^2}}{1\over g^5}+\lr{\frac{1575}{65536
   z^6}+\frac{3
   A_2^2}{128 z^4}+\frac{31 A_3}{256 z^4}-\frac{A_5}{8 z^2}}{1\over g^6} 
   +O\left(1\over g^7\right)\,.
\end{align}
Notice that the expansion coefficients contain only even powers of $1/z$.  
We would like to emphasize that the expansion 
\re{AB} is well-defined for $g\gg 1$ and $z=O(g^0)$.

Substituting \re{AB} into \re{u-der1}, we can expand $ \partial_g u$ in powers of $1/g$ 
\begin{align}\notag\label{AB1}
 \partial_g u &=4I_0+\frac{3 I_1}{8 g^2}-\frac{A_2 I_1}{2 g^3}-\lr{\frac{45}{512}I_2+\frac{1}{2} A_3
   I_1}{1\over g^4}+\lr{\frac{15 A_2
  }{64}  I_2-\frac{A_4}{2} I_1}{1\over g^5}
 \\ & 
   +\lr{\frac{1575
  }{16384} I_3+\frac{3A_2^2}{32}  
   I_2+\frac{31 A_3}{64}I_2-\frac{A_5 }{2}I_1}{1\over g^6}
   +O(1/g^{7}) 
   \, ,
\end{align}
with a notation $I_n$ introduced for the integral
\begin{align}\notag\label{profile}
I_n(y,\xi) &= {1\over\pi} \int_0^\infty  {dz\over z^{2n}} \, z \partial_z \log(1-\widehat \chi(z) )
\\
&= {1\over\pi} \int_0^\infty  {dz\over z^{2n}} \, z \partial_z  \log\lr{ { \cosh (\sqrt{z^2+\xi^2})-\cosh\xi \over \cosh (\sqrt{z^2+\xi^2})+\cosh y}}\,.
\end{align}
As we will see in a moment, the dependence of the octagon on the kinematical variables $y$ and $\xi$ enters through $I_n(y,\xi)$. 
This is the reason why we will refer to $I_n(y,\xi)$ as a profile function. Strictly speaking, the integral in \re{profile} diverges at the lower limit for $n\ge 1$ and requires a regularization. 
We address this issue in Section~\ref{sect:reg}.

\subsubsection{Quantization conditions}

The leading term in the expansion \re{AB1} looks as
\begin{align}
 \partial_g u &= 4I_0+ O(1/g^2) = 2A_0+ O(1/g^2)\,,
\end{align}
where we integrated by parts and matched the result to Eq.\ \re{A-1}. We verify that this relation is in agreement with \re{u-str}. 

Then, we replace $u$ on the left-hand side of \re{AB1} with its general 
expression \re{u-str} and compare the coefficients in front of powers of $1/g$. This gives recurrence relations between the coefficients $A_k$
\begin{align}
& 
A_0 = 2 I_0 \,, 
&&
A_2 = -{3\over 4 } I_1\,, 
&& 
A_3={3\over 4 }A_2 I_1\,, 
&&
A_4={15\over 128} I_2 +\frac23 A_3 I_1\,, \quad \dots
\end{align}
These relations allow us to express all the coefficients $A_k$ in terms of the functions $I_n(y,\xi)$ defined in \re{AB1}.
The explicit expressions for the first few of them are
\begin{align}\label{u's}\notag
&  A_0=  {2  I_0} \,,
\\[2.5mm] \notag &
A_1= 1\,,
\\[1.2mm] \notag &
   A_2= -\frac{3  I_1}{4 }, 
\\ \notag & 
   A_3=
   -\frac{9  I_1^2}{16  }\,, 
\\ \notag & 
   A_4= -\frac{3
    I_1^3}{8 }+ \frac{15  I_2}{128 }\,, 
  \\ \notag & 
   A_5=  -\frac{15  I_1^4}{64 }+\frac{75  I_1  I_2}{256 }\,, 
  \\ \notag & 
   A_6= -\frac{9  I_1^5}{64 }+\frac{225  I_2  I_1^2}{512 }-\frac{945  I_3}{8192 }\,, 
  \\ \notag & 
   A_7= -\frac{21  I_1^6}{256 }+\frac{525  I_2
    I_1^3}{1024 }-\frac{6615  I_3  I_1}{16384 }-\frac{1785  I_2^2}{16384 }\,, 
  \\ \notag & 
   A_8= -\frac{3  I_1^7}{64 }+\frac{525  I_2  I_1^4}{1024 }-\frac{6615  I_3
    I_1^2}{8192 }-\frac{1785  I_2^2  I_1}{4096}+\frac{70875  I_4}{262144}\,, 
  \\  & 
   A_9= -\frac{27  I_1^8}{1024 }+\frac{945  I_2  I_1^5}{2048 }-\frac{19845  I_3
    I_1^3}{16384 }-\frac{16065  I_2^2  I_1^2}{16384}+\frac{637875  I_4  I_1}{524288 }+\frac{292005  I_2
    I_3}{524288 }\,,
\end{align}
where we also included for completeness the values of the lowest coefficients $A_0$ and $A_1$.

It is straightforward to extend these relations to arbitrary order in $1/g$ expansion. We present the explicit expressions for the coefficients $A_k$ (with $0\le k\le 40$)  in an ancillary file with our paper.
Relations \re{u's} combined with \re{O-str}  yield a strong coupling expansion of the octagon.

We notice that $A_3=-A_2^2$. Similar relations hold for all coefficients with odd indices. Namely, the coefficients $A_{2k+1}$ can be expressed in terms of the even coefficients 
$A_{2m}$ (with $m\le k$), e.g.,
\begin{align}\notag\label{u-odd}
& A_3=-A_2^2\,,
\\[1.5mm]\notag
& A_5=\frac{20 A_2^4}{9}-\frac{10 A_2 A_4}{3}\,,
\\\notag
& A_7=-\frac{112
   A_2^6}{5}+\frac{1568}{45} A_4 A_2^3-\frac{14 A_6A_2}{3}-\frac{119
   A_4^2}{15}\,,
\\
& A_9=\frac{68032 A_2^8}{135}-\frac{42976}{45} A_4 A_2^5+\frac{432}{5} A_6
   A_2^3+\frac{2008}{5} A_4^2 A_2^2-6 A_8 A_2-\frac{206 A_4 A_6}{5}\,.
\end{align}
The origin of these relations is elucidated in Appendix~\ref{app:sol}.

\subsubsection{Profile function}\label{sect:reg}

A close examination shows that the integral in \re{profile} is not well-defined for positive integer $n$ and requires a regularization. Indeed, it is easy to see that $\log(1-\widehat \chi(z) )\sim \log z$ for $z\to 0$ 
and, therefore, the integral in \re{profile} diverges at the lower limit as $\int  dz/z^{2n}$. 

To understand the origin of this divergence, we turn back to Eq.\ \re{u-der1}. The integral in \re{u-der1} is convergent at small $z$ and the divergences appear in \re{AB1} after we exchanged the integration and 
series expansion of the integrand at large $g$. As mentioned above, the series expansion is well-defined for $z=O(g^0)$ and, therefore, performing the integration we should have imposed a lower cut-off 
on the possible values of $z$. 

The simplest way to do it is by introducing the so-called analytical regularization (see Appendix~\ref{app:reg}).  Namely, we modify the definition \re{profile} by inserting an addition factor $z^\epsilon$ that 
suppresses the contribution of small $z$
\begin{align}\label{I-anal}
I_{n}(\epsilon) &= {1\over\pi} \int_0^\infty  {dz\,z^{\epsilon}\over z^{2n}} \, z \partial_z \log(1-\widehat \chi(z) )\,.
\end{align}
For any given $n$ we choose $\epsilon$ sufficiently large so that the integral is convergent. Then, integrating by parts, we can reduce the strength of
singularity at small $z$ 
\begin{align}
I_n(\epsilon)  ={1\over (2n-\epsilon-1)\pi} \int_0^\infty  {dz\,z^{\epsilon}\over z^{2n-2}} \times {1\over z}\partial_z  z \partial_z \log(1-\widehat \chi(z) )\,.
\end{align}
 Subsequently integrating by parts $n$ times we arrive at the integral that is well-defined for $\epsilon\to 0$. Taking the limit $I_n= \lim_{\epsilon\to 0} I_n(\epsilon)$ we arrive at 
\begin{align} \notag\label{profile1}
I_n  
 &={1\over (2n-1)!! \pi} \int_0^\infty  {dz} \, \big(z^{-1}\partial_z\big)^n    z \partial_z \log(1-\widehat \chi(z) )
 \\
&= {1\over (2n-1)!! \pi} \int_0^\infty  {dz} \, \big(z^{-1}\partial_z\big)^n    z \partial_z  \log\lr{ { \cosh (\sqrt{z^2+\xi^2})-\cosh\xi \over \cosh (\sqrt{z^2+\xi^2})+\cosh y}}\,.
\end{align}
It is easy to verify using  $z \partial_z \log(1-\widehat \chi(z) )\sim 1+ O(z^2)$ that it is well-defined for  $z\to 0$.
The explicit expressions for the profile function $I_n(y,\xi)$ in different kinematical limits are given in the next section.

\section{Properties of strong coupling expansion}\label{sect:str}

At strong coupling, the octagon is given by \re{O-str} with the expansion coefficients defined in \re{u's}. In this section, we examine properties of the obtained expressions. 

\subsection{Improved expansion}

According to \re{u's}, the expansion coefficients $A_n$ (with $n\ge 2$) are given by multi-linear combinations of the profile functions $I_k$ 
\begin{align}
A_{n} = c_n I_1^{n-1} +\sum_{\ell\ge 2} \sum_{ p_1,\dots, p_\ell \ge 0}  c_{p_1,\dots,p_\ell} \, I_1^{p_1} \dots  I_\ell^{p_\ell}\,,
\end{align}
where the nonnegative integers $p_i$ satisfy the equation
\begin{align}
 p_1 +3p_2 + \dots + (2\ell-1)p_\ell = n-1 \,.
\end{align}

The rational coefficients $c_n$ and $c_{p_1,\dots,p_\ell}$ exhibit a remarkable regularity. 
To make this manifest, we rewrite the octagon \re{O-str} in the following form
\begin{align}\label{O-str1}
\log \mathbb O = - 2 g I_0 + \frac18  \, \log g + B +\log \mathbb O_{\rm q}\,,
\end{align}
where $\log \mathbb O_{\rm q}$ is given by 
\begin{align}\label{O-I1}
\log \mathbb O_q=\frac38  \, \log g+ {A_2\over 4g} + {A_3\over 12 g^2}+ {A_4\over 24 g^2} + \dots
\end{align}
Replacing the coefficients $A_n$ with their expressions \re{u's}, we find 
\begin{align} \label{Oq-exp}
& \log \mathbb O_q =\frac38  \, \log g -\frac{3 I_1}{16
   g}-\frac{3 I_1^2}{64 g^2}+\lr{\frac{5
   I_2}{1024}-\frac{I_1^3}{64}}{1\over g^3}+\lr{\frac{15 I_1
   I_2}{2048}-\frac{3 I_1^4}{512}}{1\over g^4}
\\[2mm]   \notag
  & +\lr{-\frac{3
   I_1^5}{1280}+\frac{15 I_2 I_1^2}{2048}-\frac{63
   I_3}{32768}}{1\over g^5}+\lr{-\frac{I_1^6}{1024}+\frac{25 I_2
   I_1^3}{4096}-\frac{315 I_3 I_1}{65536}-\frac{85
   I_2^2}{65536}}{1\over g^6}+
  \dots
\end{align}
Notice that the coefficients involve terms of the form $I_1^\ell$, $I_2 I_1^\ell, I_3 I_1^\ell, \dots$. Such terms can be resummed to all orders in $\ell$
\begin{align}\label{g'}
\log {\mathbb O}_{\rm q} 
=\frac{3}{8} \log (g-I_1/2)+\frac{5 I_2}{1024 (g-I_1/2)^3}-\frac{63 I_3}{32768 (g-I_1/2)^5}
+\dots\,,
\end{align}
where the dots denote further terms of the infinite series. 

The relation \re{g'} suggests to change the expansion parameter to $g'=g-I_1/2$. Expanding $\log {\mathbb O}_{\rm q} $ at large $g'$, we get a remarkably simple expression
\begin{align}\label{Oq}
\log  {\mathbb O}_{\rm q} =\frac{3 \log (g')}{8}+\frac{5 I_2}{1024 g'{}^3}-\frac{63 I_3}{32768
   g'{}^5}-\frac{85 I_2^2}{65536 g'{}^6}+\frac{10125 I_4}{4194304
   g'{}^7}+\frac{32445 I_2 I_3}{8388608 g'{}^8}+O\left(1/ g'{}^9\right),
\end{align}
where the expansion coefficients do not depend on $I_1$. The expressions for the octagon \re{O-str1} and \re{Oq} up to order $O(1/g^{38})$ can be found in an ancillary file. 

The situation here is similar to that of the strong coupling expansion of the cusp anomalous dimension. In the latter case, all 
corrections proportional to $\log 2$ could be absorbed into the redefinition of the coupling constant~\cite{Basso:2007wd}. In the present case, the shifted   coupling constant, $g'=g-I_1/2$, depends on the  
kinematical variables $y$ and $\xi$.
 
\subsection{Kinematical limits} 

In this subsection, we evaluate the profile function \re{profile1} and study the properties of the octagon at strong coupling in different kinematical regimes.

\subsubsection*{Symmetric point $\boldsymbol{y=\xi=0}$}

In this kinematical point, the profile function \re{profile1} is given by $I_0=\pi/2$ and $I_1=-2\log 2/\pi$. For $n\ge2$, we have
\begin{align} \label{prof1}
I_n 
& =(-1)^n \left(1-2^{2-2n}\right) {2\zeta (2 n-1)\over \pi^{2n-1}} = (-1)^n {2\eta(2n-1)\over \pi^{2n-1}}\,.
\end{align}
This relation should be compared with its counterpart at weak coupling \re{q-weak00}.

Substituting \re{prof1} into \re{u's}, we obtain for the octagon \re{O-str}
\begin{align}\notag\label{oct(00)}
\log \mathbb O& =-\pi  g+\frac{1}{2}\log g+B+ \frac{3 \log (2)}{8 \pi  g}-\frac{3 \log ^2(2)}{16 \pi ^2
   g^2}+\lr{\frac{15 \zeta (3)}{2048 \pi ^3}+\frac{\log ^3(2)}{8 \pi
   ^3}}{1\over g^3}
   \\\notag
&   - \lr{\frac{45 \zeta (3) \log (2)}{2048 \pi ^4}+\frac{3 \log ^4(2)}{32 \pi
   ^4}}{1\over g^4}+\lr{\frac{945 \zeta (5)}{262144 \pi ^5}+\frac{45 \zeta (3) \log
   ^2(2)}{1024 \pi ^5}+\frac{3 \log ^5(2)}{40 \pi ^5}}{1\over g^5}
\\ &   
   -\lr{\frac{765 \zeta
   (3)^2}{262144 \pi ^6}+\frac{75 \zeta (3) \log ^3(2)}{1024 \pi ^6}+\frac{4725 \zeta (5)
   \log (2)}{262144 \pi ^6}+\frac{\log ^6(2)}{16 \pi
   ^6}}{1\over g^6} + O(1/g^7)
   \, .
\end{align}
As explained earlier, all terms in this expression involving  $\log(2)$  can be absorbed into the redefinition of the coupling $g'=g+\log 2/\pi$.
Assigning a weight $1$ to $\pi$ and $\log(2)$ and weight $k$ to $\zeta(k)$, 
we observe that the first two terms in \re{oct(00)} possess weight $1$, whereas all terms suppressed by powers of $1/g$ have a uniform weight $0$. 

A close examination of \re{oct(00)} 
shows that, starting from the $O(1/g)$ term, 
the expansion coefficients have alternating signs and grow factorially at higher orders.~\footnote{To check 
this, we computed the coefficients up to order $O(1/g^{38})$, see \re{bulk-num} below.} This suggests that the strong coupling expansion \re{oct(00)} is Borel summable.

As was mentioned above, $B$ in \re{oct(00)} arises as an integration constant in \re{D-u} and, therefore, it does not depend on the coupling constant. For $y=\xi=0$, we can fix its value by comparing 
\re{oct(00)} with the numerical result for the octagon at some reference value of the coupling $1< g <10$. In this manner, we deduce that $B=0.960877$\,.

\subsubsection*{Single-trace OPE channel $\boldsymbol{\phi=0, \ \xi\gg 1}$}

In this limit, for $y=i\pi$ and large $\xi$, the cut-off function \re{chi} simplifies to $\widehat \chi(z) =\exp(-z^2/(2\xi))$. 
Changing the integration variable in  \re{profile1} to $x=z/\sqrt{\xi}$, we get  for $\xi\gg 1$
\begin{align}  
I_n 
 & = {\xi^{1/2-n}\over (2n-3)!!} {(-1)^{n+1}\over \sqrt{2\pi}} \zeta\lr{\ft32-n} 
 =2^{\frac{3}{2}-n} {\zeta \left(n-\frac{1}{2}\right)\over (2\pi \xi)^{n-1/2}} \sin \left(\ft{\pi}{4}(2n-1) \right)\,.
\end{align}
This relation holds up to corrections suppressed by powers of $1/\xi$.

The strong coupling expansion of the octagon, Eqs.~\re{O-str} and \re{u's}, takes the form
\begin{align}\notag\label{oct(01)}
& \log \mathbb O = -\frac{  \zeta \left(\frac{3}{2}\right)}{\pi }({2\pi \xi})^{1/2}g+\frac{\log
   g}{2}+B
   -\frac{3 \zeta\left(\frac{1}{2}\right)}{16} {1\over \lr{2\pi \xi}^{1/2} g}
   -\frac{3 \zeta \left(\frac{1}{2}\right)^2}{64}{1\over 2\pi \xi g^2}
 \\   \notag &
  + \left(\frac{5 \zeta
   \left(\frac{3}{2}\right)}{2048}-\frac{\zeta
   \left(\frac{1}{2}\right)^3}{64}\right)\frac{1}{\lr{2\pi \xi}^{3/2}g^3}   
   -\lr{\frac{3 \zeta
   \left(\frac{1}{2}\right)^4}{512}-\frac{15 \zeta \left(\frac{1}{2}\right)
   \zeta \left(\frac{3}{2}\right)}{4096}}{1\over {(2\pi \xi)^2 g^4}}
  \\ &  \notag
   +\left(-\frac{3 \zeta \left(\frac{1}{2}\right)^5}{1280}+\frac{15
   \zeta \left(\frac{1}{2}\right)^2 \zeta \left(\frac{3}{2}\right)}{4096}+\frac{63 \zeta
   \left(\frac{5}{2}\right)}{131072}\right) \frac{1}{\lr{2\pi \xi}^{5/2}g^5} 
  \\ & 
   -\lr{\frac{\zeta
   \left(\frac{1}{2}\right)^6}{1024}-\frac{25 \zeta \left(\frac{1}{2}\right)^3 \zeta
   \left(\frac{3}{2}\right)}{8192}+\frac{85 \zeta
   \left(\frac{3}{2}\right)^2}{262144}-\frac{315 \zeta \left(\frac{1}{2}\right) \zeta
   \left(\frac{5}{2}\right)}{262144}} 
   {1\over (2\pi\xi)^3 g^6}
   +O\left(1/g^7\right)\,.
\end{align}
The expansion coefficients in this relation involve $\zeta-$functions evaluated at half-integer positive values. 
Applying \re{Oq}, we can eliminate $\zeta(\ft12)=-1.46035$ from \re{oct(01)} by redefining the coupling constant $g'=g- {\zeta
   \left(\frac{1}{2}\right)}/(8 \pi  \xi)^{1/2}$.
   
The series \re{oct(01)} is very similar to \re{oct(00)} as far as the properties of the expansion coefficients are concerned. Namely, starting from the $O(1/g)$ term, the coefficients have a uniform weight $0$.
In addition, they are sign-alternating and grow factorially at higher orders.    
   
We observe that in all terms in Eq.\ \re{oct(01)}, except the second one, the coupling constant is accompanied by $(2\pi \xi)^{1/2}$, so that 
the expansion parameter is effectively $(2\pi \xi)^{1/2} g$. Assuming that the $O(g^0)$ term in \re{oct(01)} has the same property, we can predict the 
$\xi-$dependence of the constant term $B$
\begin{align}
B= \frac14 \log(2\pi\xi) + c\,.
\end{align}
We verified that this relation agrees with the numerical values of the octagon at large $\xi$ and extracted the constant $c=0.343754$.

\begin{figure} 
\psfrag{g1}{$\tilde g$}
\psfrag{O}[cr][cc]{$\log \mathbb O\ \ $}
\centerline{\parbox[c]{82mm}{\insertfig{8.2}{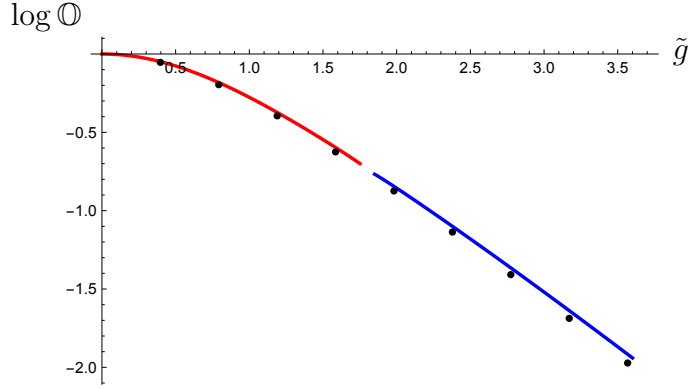}}}
\caption{Dependence of $\log \mathbb O$ on $\tilde g=(2\pi \xi)^{1/2} g$ at weak (red curve) and strong (blue curve) coupling for $\xi\to\infty$. Black dots are numerical values of  $\log \mathbb O$ 
at $\xi=10$ and different $\tilde g$. They lie slightly below the curves, the difference is due to $O(1/\xi)$ correction. }\label{fig:large-xi}
\end{figure}

It is interesting to note that the  weak coupling expansion of the octagon  \re{weak-eff} also runs in (even) powers of $(2\pi \xi)^{1/2} g$. This suggests to introduce $\tilde g = (2\pi \xi)^{1/2} g$ and 
compare the dependence of the octagon on $\tilde g$ at weak and strong coupling. Neglecting corrections suppressed by powers of $1/\xi$, we observe (see Figure~\ref{fig:large-xi}) that the two 
curves defined by \re{weak-eff}  and \re{oct(01)} merge at intermediate values of $\tilde g$.

\subsubsection*{Double-trace OPE channel $\boldsymbol \xi=0$ and $\boldsymbol {\phi\to  0}$} 
 
As was already emphasized in Section~\ref{sect:lim}, the leading contribution to the correlation function in this limit comes from double-trace operators. We expect that for $\phi\to 0$ with $g$ kept fixed, 
the octagon should take a finite value $\mathbb O\to 1$.

To approach this limit, we replace $y=i(\pi-\phi)$ and $\xi=0$ in the profile function \re{profile1} and take $\phi\to 0$.
The leading contribution to the integral in \re{profile1} comes from $z=O(\phi)$ whereas  integration over $z>1$ in \re{profile1} yields a subleading contribution as $\phi\to 0$. Replacing 
the integration variable to $x=z/\phi$ with $0\le x\le 1/\phi$, we can expand the integrand in \re{profile1} in powers of $\phi^2$ using the relation
\begin{align}
 z \partial_z\log\lr{1-\widehat\chi(z)} = x \partial_x \log\lr {\cosh (x\phi)-1 \over \cosh (x\phi)-\cos \phi}  = {2\over 1+x^2} - \phi^4 {x^2\over 120} + O(\phi^6)\,.
\end{align}
This leads to the following result for the profile function $I_n$ for $n\ge 1$
\begin{align}\label{prof-sing}
I_n =  {\phi^{1-2n}\over (2n-1)!! \pi} \int_0^\infty  {dx} \, \big(x^{-1}\partial_x\big)^n {2\over 1+x^2} + O(\phi) = (-1)^n \phi^{1-2n} + O(\phi)\,,
\end{align}
where the last term accounts for the contribution from $z=O(\phi^0)$. For $n=0$ we have $I_0=\phi -\phi^2/(2\pi)$.

Substituting \re{prof-sing} into \re{O-str} and \re{u's}, we obtain the strong coupling expansion of the octagon
\begin{align}\notag\label{phi-exp}
\log \mathbb O = {}&    -2 g \phi +B+\frac{1}{2}\log g+\frac{3}{16 g \phi }-\frac{3}{64
   g^2 \phi ^2}+\frac{21}{1024 g^3 \phi ^3}
   \\ &
   -\frac{27}{2048 g^4 \phi
   ^4}+\frac{1899}{163840 g^5 \phi ^5}-\frac{27}{2048 g^6 \phi
   ^6}+\dots\,,
\end{align}
where the dots stand for higher-order terms in $1/g$ as well as terms vanishing as $\phi\to 0$.

Notice that the expansion in \re{phi-exp} runs in powers of $1/(g\phi)$ and, in order to take the limit $\phi\to 0$, the series needs to be resummed. It is convenient to switch to $\alpha=8g\phi$  and examine 
the limit $\phi \to 0$ with $\alpha$ held fixed and small. Then, the relation \re{phi-exp} takes the form
\begin{align} \label{ser-alpha}
 \log \mathbb O =   -\frac{\alpha}{4}+ B+\frac{\log
   (g)}{2} +\frac{3}{2\alpha} -{3\over\alpha^2}+\frac{21}{2\alpha^3}-{54\over\alpha^4}+\frac{1899}{5\alpha^5}-{3456\over
   \alpha^6}+O\left(1/\alpha^{7}\right) \,.
\end{align}
Assuming that $\log \mathbb O$ depends on $\phi$ and $g$ through $\alpha$, we can determine the asymptotic behavior of $B$ at small $\phi$
\begin{align}\label{B-res}
B= \frac12 \log (\pi\phi)\,.
\end{align}
In general, we can add an arbitrary constant to the right-hand side of this relation. As we show in a moment, the condition
for $\mathbb O$ to vanish for $\phi\to 0$ leaves however no room for it.
 
\begin{figure} 
\psfrag{alpha}{$\alpha$}
\psfrag{O}[cr][cc]{$\log \mathbb O$}
\centerline{\parbox[c]{82mm}{\insertfig{8.2}{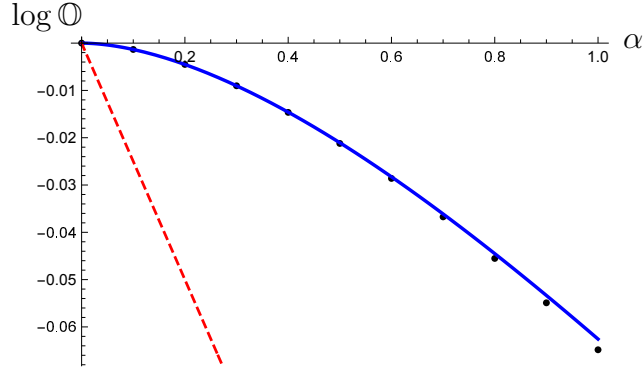}}}
\caption{Dependence of the octagon on $\alpha=8g\phi$ for $g=10$ and $\xi=0$. Blue line shows \re{O-resumed}, red dashed line describes the leading strong coupling result (the first term in \re{ser-alpha}). Black dots are numerical values of $\log \mathbb O$ computed using \re{oct} with the size of the matrix $k_-$ truncated to $N_{\rm max}=100$.}
\label{fig:dt}
\end{figure}

To get a closed expression for the series \re{ser-alpha}, we examine the expansion coefficients of  $\partial_\alpha {\log \mathbb O}$. It turns out that they form a sequence that had already appeared in the 
literature, see Refs.~\cite{martin2011exactly,OEIS}, leading to 
\begin{align}\notag\label{der-O}
  \partial_\alpha {\log \mathbb O} &=-\sum_{n\ge 0} {(-1)^n\over \alpha^{n}} \int_0^\infty {ds \, s^{n-2}\over  K_1^2(s/4) + \pi^2 I_1^2(s/4) }
\\
& = - \int_0^\infty {ds  \over s^2 (K_1^2(s/4) + \pi^2 I_1^2(s/4)) }{\alpha \over (s+\alpha)}\,,
\end{align}
where $K_1$ and $I_1$ are modified Bessel functions. The integrand decreases exponentially fast at large $s$  and behaves as $\alpha/(16(\alpha+s))$ at small $s$. 
Integrating by parts, we get for small $\alpha$
\begin{align} \label{dO}
&  \partial_\alpha {\log \mathbb O} =  \frac{\alpha}{16} \left( \log \alpha - \kappa + O(\alpha)\right)\,,
\end{align}
where 
\begin{align}
\kappa= - \int_0^\infty ds\,  {d\over ds} \left[{16 \over s^2( K_1^2(s/4) + \pi^2 I_1^2(s/4)) } \right] \log s
= 3\log 2-\gamma\,.
\end{align}
Imposing the condition that $\log \mathbb O$ has to vanish for $\alpha\to 0$, we finally arrive at
\begin{align}\label{O-resumed}
\log \mathbb O = \frac{1}{32}\alpha^2 \log \alpha - \frac{1+2\kappa}{64}\alpha^2 + O(\alpha^3)\,.
\end{align}
We recall that this relation was obtained for $\phi\to 0$ and  $\alpha=8g\phi$ fixed. 
The comparison of \re{O-resumed} with the numerical value of $\log \mathbb O$ at $g=10$ is shown in Figure~\ref{fig:dt}. 
Integrating \re{der-O} with the boundary condition \re{O-resumed}, we can obtain $\log\mathbb O$ for arbitrary $\alpha$. Its 
expansion at large $\alpha$ takes the form \re{ser-alpha} with $B$ given by \re{B-res}. 

Notice that the resummed expression \re{O-resumed} vanishes for $\alpha\to 0$ faster than  the leading contribution 
at strong coupling.  The latter is described by the first term on the right-hand side of \re{ser-alpha}. 
It is also interesting to note that the leading asymptotic behavior of the octagon at strong coupling $\log \mathbb O \sim 2 g^2 \phi^2 \log(g \phi)$ is remarkably similar to that at weak coupling $\log \mathbb O 
\sim 2 g^2 \phi^2 \log\phi$ (see Eq.~\re{log-phi}) even though the two expressions are valid in different regions of the parameter space.
 
\subsubsection*{Null limit $\boldsymbol{y\gg 1, \ \xi< 1}$} 

In this limit, the profile function \re{profile1} can be expanded in powers of $\xi^2$
\begin{align}\label{I-exp}
I_n=I_n^{(0)} + \xi^2 I_n^{(1)} + \xi^4 I_n^{(2)} + O(\xi^6)\,,
\end{align}
where the coefficient functions $ I_n^{(i)}(y)$ are computed in Appendix~\ref{app:prof}.

Using \re{I-exp}, we derive the strong coupling expansion of the octagon \re{O-str} and \re{u's}
\begin{align}\label{O-sum-f}
\log \mathbb O = f^{(0)}(y) + \xi^2   f^{(1)}(y) + \xi^4 f^{(2)}(y) + O(\xi^6)\,.
\end{align}
The leading term is  
\begin{align}\notag\label{O-0}
f^{(0)} & =-
g \left(\frac{y^2}{\pi }+\pi
   \right)+\frac{\log g}{2}+B{}^{(0)} -\frac{3 L}{16 \pi
    g}-\frac{3 L^2}{64 \pi ^2
   g^2}-\lr{\frac{L^3}{64 \pi ^3}+\frac{5
   \zeta (3)}{4096 \pi ^3}}{1\over g^3}
\\ 
&   -\lr{\frac{3
   L^4}{512 \pi ^4}+\frac{15 L \zeta (3)}{8192
   \pi ^4}}{1\over g^4}-\lr{\frac{3 L^5}{1280 \pi
   ^5}+\frac{15 L^2 \zeta (3)}{8192 \pi
   ^5}+\frac{63 \zeta (5)}{524288 \pi
   ^5}}{1\over g^5} +O\left(1\over g^6\right)\,,
\end{align}
with the notation introduced for $L=\log y +\gamma-\log(2\pi)$. 
The subleading terms are given by
\begin{align}
\notag\label{O-1}
f ^{(1)}
& = 
g  \frac{L+1}{\pi } +B{}^{(1)}+ \frac{9 \zeta (3)}{128 \pi ^3
   g}+\frac{9 L \zeta (3)}{256 \pi ^4
   g^2}+\lr{\frac{9 L^2 \zeta (3)}{512 \pi
   ^5}+\frac{75 \zeta (5)}{32768 \pi
   ^5}}{1\over g^3}
 \\  
&   +\lr{\frac{9 L^3 \zeta (3)}{1024
   \pi ^6}+\frac{225 L \zeta (5)}{65536 \pi
   ^6}+\frac{45 \zeta (3)^2}{65536 \pi
   ^6}}{1\over g^4}+ O\left(1\over g^5\right)\,,
\\[2mm]
\notag\label{O-2}
f ^{(2)}
&=-g\frac{5 \zeta (3)}{16 \pi ^3}+B{}^{(2)}-\frac{105 \zeta (5)}{2048 \pi ^5 g}-\lr{\frac{105 L
   \zeta (5)}{4096 \pi ^6}+\frac{27 \zeta (3)^2}{4096 \pi ^6}}{1\over g^2}
\\ &  
   -\lr{\frac{105 L^2
   \zeta (5)}{8192 \pi ^7}+\frac{27 L \zeta (3)^2}{4096 \pi ^7}+\frac{1575 \zeta
   (7)}{524288 \pi ^7}}{1\over g^3} + O\left(1\over g^4\right)\,.
\end{align}
Here $B{}^{(i)}$ are the expansion coefficients of $B$ in powers of $\xi^2$. Numerical analysis shows that
\begin{align}
B = \lr{0.035 \, y^2 +0.313 \, L  + 1.208}  + \xi^2 (-0.171\, L +0.127 ) + O(\xi^4) \,.
\end{align}
As before, the expansion coefficients in front of powers of $1/g$ in the expressions for $f^{(i)}$ have a uniform weight $0$. 

The expressions for $f^{(i)}$ contain terms enhanced by powers of $L=\log y +\gamma-\log(2\pi)$. Such terms can be resummed to all orders
using the property of the octagon exhibited in Eqs.~\re{O-str1} and \re{Oq}, leading to
\begin{align}\label{O-impr}
\log \mathbb O =  - 2 g I_0 + \frac18  \, \log g+\frac{3}{8} \log (g') + B +\frac{5 I_2}{1024 g'{}^3}-\frac{63 I_3}{32768
   g'{}^5} + O(1/ g'{}^6)\,,
\end{align}
where $g'=g-I_1/2$ and the profile functions $I_0$, $I_1$, $\dots$ are given by
\begin{align}\notag
& I_0 = \frac{y^2+\pi ^2}{2 \pi }-\frac{(L+1)}{2 \pi } \xi ^2+\frac{5\zeta (3)}{32 \pi
   ^3} \xi ^4  + O(\xi^6)\,,
\\ \notag
& I_1=\frac{L}{\pi }-\frac{3 \zeta (3)}{8 \pi ^3} \xi ^2+\frac{35 \zeta (5)}{128 \pi ^5} \xi ^4 + O(\xi^6)\,,
\\ \notag
& I_2=-\frac{\zeta (3)}{4 \pi ^3}+\frac{15  \zeta (5)}{32 \pi
   ^5}\xi ^2 -\frac{315  \zeta (7)}{512 \pi ^7}\xi ^4+ O(\xi^6)\,,
\\
&I_3=\frac{\zeta (5)}{16 \pi ^5}-\frac{35  \zeta (7)}{128 \pi
   ^7}\xi ^2+  \frac{1155 \zeta (9)}{2048 \pi ^9}\xi ^4 +O(\xi^6)\,.
\end{align}
The first term in \re{O-impr} agrees with \re{InfinitGoctagon}, as anticipated. The expansion \re{O-impr} is well-defined for large $g'=g-I_1/2$, or equivalently for $g\gg L/(2\pi)$.

Comparing \re{O-sum-f} with \re{WeakOcta}, we notice that the dependence of the null octagon on $\xi$ changes as we go from weak to strong coupling. In the former case, it is one-loop 
exact $\log \mathbb O\sim g^2 \xi^2$ whereas in the latter case, $\log \mathbb O$ scales as $O(g)$ and its small $\xi-$expansion does not truncate.

The transition between the two regimes can be understood using the last relation in \re{eqs}. We demonstrated in Section~\ref{sect:weak}, that at weak coupling the second term in the right-hand 
side of this relation is exponentially small at large $y$. At strong coupling, we replace the function $q(z)$ in \re{eqs} with its leading asymptotic behavior \re{q-gen} 
and take into account \re{Q0-mom} to get
\begin{align}\label{neg1}
\partial_\xi u & = - 8 g^2 \xi +8 g^2 \sinh  \xi   \int_0^\infty  \frac{dz\,  z \, [J_1(2 g z)]^2}{ \cosh \left(\sqrt{\xi
   ^2+z^2}\right)-\cosh (\xi )} + \dots\,,
\end{align}
where the ellipses denote corrections suppressed by $1/g$. At large $g$, the leading contribution to the integral comes from $z=O(1/g)$.
Changing the integration variable to $x=2 gz$  and expanding the integrand at large $g$, we find that the second term 
on the right-hand side of \re{neg1} behaves at small $\xi$ as
\begin{align}
8g^2 \xi \left[2 \int_0^\infty {dx\over x} J_1^2(x) + O(1/g)\right] = 8g^2\xi \left[ 1+O(1/g)\right].
\end{align}
Substituting this relation into \re{neg1}, we verify that the leading $O(g^2)$ term cancels yielding $\partial_\xi u=O( g \xi)$
or equivalently $\log \mathbb O =O( g\xi^2)$ in agreement with \re{O-sum-f}.

\section{Numerical checks}
\label{SectNumerics}

In this section, we compare numerical values of the octagon in different kinematical limits with the corresponding analytical expressions 
obtained in the previous section. The numerical results presented in this section were obtained in collaboration with Riccardo Guida.

To compute $\mathbb O$ numerically for finite 't Hooft coupling, we apply the relations \re{oct} and \re{oct1} and truncate the size of the matrix $k_-$ to $N_{\rm max}=100$. In addition, 
we also compute a logarithmic derivative of the octagon with respect to the coupling constant, $\partial_g \log\mathbb O$. According to \re{D-u}, it is related to the potential, 
$u= -2g\partial_g \log\mathbb O$, which can be found from \re{u-rep} by replacing  the matrix $k_-$ with its finite-dimensional minor.

\subsection{Borel--Pad\'e improvement}

At strong coupling, the octagon is given by the series expansion in $1/g$, see Eqs.~\re{O-str1} -- \re{Oq-exp}. To study its asymptotic properties, it is advantageous to consider the 
logarithmic derivative $\partial_g \log\mathbb O$. It does not contain the constant term $B$ and its expansion  involves only inverse powers of the coupling.

To begin with, we consider  the octagon at the symmetric point $y=\xi=0$.
 According to \re{oct(00)}, it is given by a sign-alternating series in $1/g$ with factorially growing coefficients. Replacing $\zeta-$values by their numerical values, we obtain 
\begin{align}\notag\label{bulk-num}
&  \partial_g \log \mathbb O =
-3.14159+\frac{0.5}{g}-\frac{0.08274}{g^2}+\frac{0.01826}{g^3}-\frac{0.00488}{g^4}+\frac{0.00164}{g^5}-\frac{0.00067}{g^6}+\frac{0.00033}{g^7}
 \\ &\notag
-\frac{0.000194}{g^8}+\frac{0.000131}{g^9}
   -\frac{0.000099}{g^{10}}+\frac{0.000084}{g^{11}}
   -\frac{0.000078}{g^{12}}+\frac{0.000079}{g^{13}}-\frac{0.000086}{g^{14}}+\frac{0.00010}{g^{15}}
\\ &   \notag   
    -\frac{0.00013}{g^{16}}
   +\frac{0.
   00017}{g^{17}}-\frac{0.00024}{g^{18}}+\frac{0.00036}{g^{19}}-\frac{0.00057}{g^{20}}+\frac{0.00095}{g^{21}}
   -\frac{0.00165}{g^{22}}+\frac{0.00301}{g^{23}}
 \\ &   \notag   
   -\frac{0.00572}{g^{24}
   }+\frac{0.01132}{g^{25}}-\frac{0.02334}{g^{26}}+\frac{0.04995}{g^{27}}
   -\frac{0.11091}{g^{28}}+\frac{0.25514}{g^{29}}    -\frac{0.60729}{g^{30}} +\frac{1.49399}{g^{31}} 
\\ &     
   -\frac{3.79458}{g^{32}} +\frac{9.
   9407}{g^{33}}-\frac{26.8349}{g^{34}} 
  +\frac{74.5819}{g^{35}} -\frac{213.233}{g^{36}}+\frac{626.651}{g^{37}}-\frac{1891.58}{g^{38}}+\frac{5860.64}{g^{39}}+\dots
\end{align}
Following the standard procedure \cite{Bender}, we can improve the strong coupling expansion \re{bulk-num} by applying the Borel transformation
\begin{align}
 \partial_g \log \mathbb O = \int_0^\infty dt \, {\mathcal B}(t/g) \e^{-t}\,,
\end{align}
and replacing a partial sum ${\mathcal B}(t)$ by its Pad\'e approximant $[n/m]= P_n(t)/Q_m(t)$. 

\begin{figure}[t!] 
\psfrag{g}{$g$}
\psfrag{dO}[cr][cc]{$\partial_g \log \mathbb O $}
\centerline{\parbox[c]{87mm}{\insertfig{8.7}{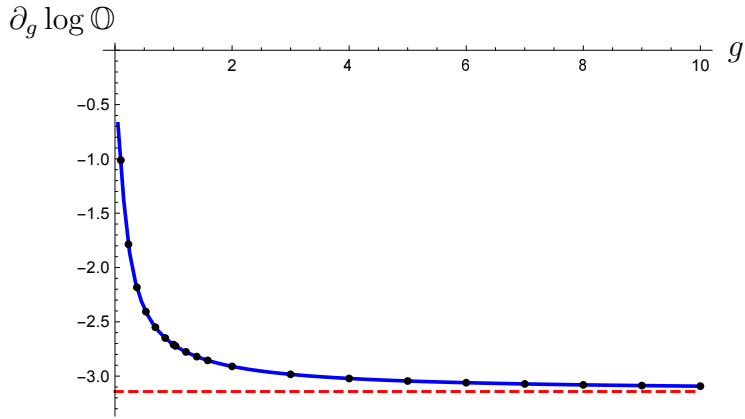}}}
\caption{Dependence of $\partial_g \log\mathbb O$ on the coupling constant for $y=\xi=0$. Solid line shows the strong coupling expansion \re{bulk-num} improved using the 
Borel-Pad\'e method, red dashed line describes the leading strong coupling result. Black dots denote numerical values.}
\label{fig:bulk}
\end{figure}  

For $n=2$ and $m=3$, the resulting expression for $\partial_g \log \mathbb O$ is shown in Figure~\ref{fig:bulk}. It agrees with the numerical values 
of the octagon up to $g=0.1$.  

We encounter similar situation in the single-trace regime, for $\phi=0$ and $\xi\gg 1$. Strong coupling expansion of the octagon \re{oct(01)} is Borel summable and it can be improved 
using the Borel--Pad\'e method. We verified that for $\xi=4$ the resulting strong coupling expansion agrees with numerical values of $\partial_g \log \mathbb O$ up to $g=0.1$.

\subsection{Order of limits phenomenon} 

As mentioned in the Introduction, in the null limit, for $y\gg 1$, the octagon has different asymptotic behavior  depending on the hierarchy between $y$ and $g$.
At weak coupling, for $y\gg g$, the octagon scales as \re{WeakOcta}. At strong coupling, for $g\gg y\gg 1$, its leading behavior is given by \re{InfinitGoctagon} and 
subleading corrections are defined in Eqs.\ \re{O-sum-f} and \re{O-impr}.

\begin{figure}[t!] 
\psfrag{g}{$g$}
\psfrag{dO}[cc][cc]{$\partial_g \log \mathbb O(0)$}
\centerline{\parbox[c]{82mm}{\insertfig{8.2}{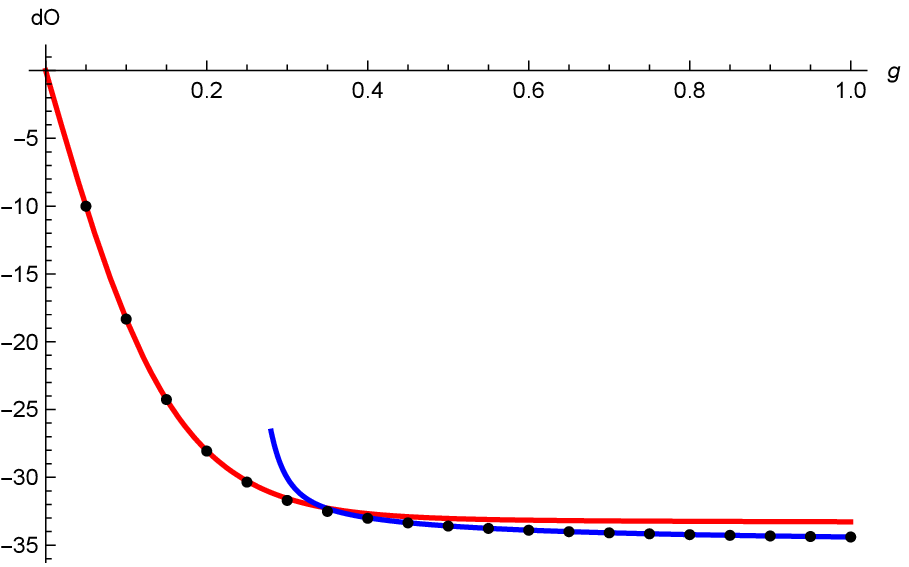}}  
\qquad
\psfrag{dO}[cc][cc]{$\partial_g \log {\mathbb O(0)\over \mathbb O(\xi))}$} 
\parbox[c]{82mm}{\insertfig{8.2}{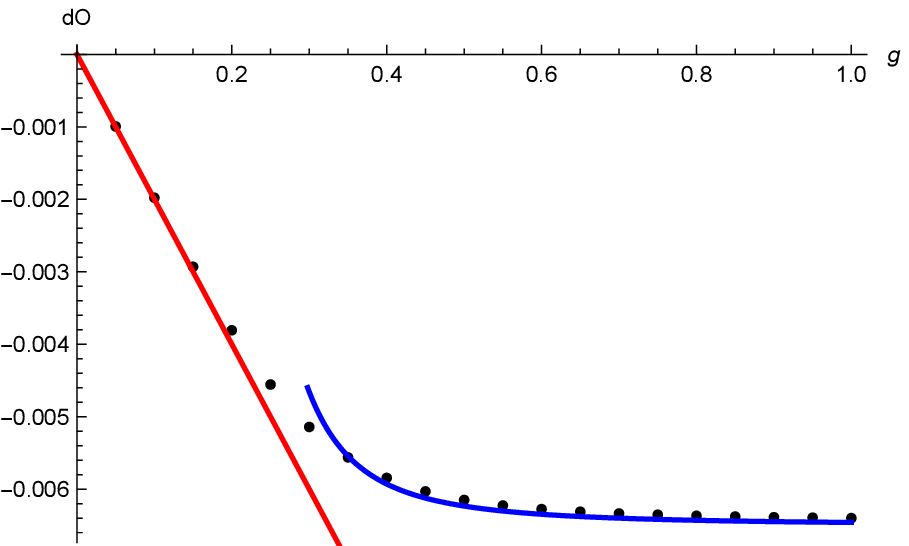}}}
\caption{Dependence of the octagon on the coupling constant at $y=10$ for $\xi=0$ (left panel) and $\xi=1/10$ (right panel). 
Black dots denote numerical values. Red and black lines depict \re{WeakOcta} and \re{O-impr}, respectively. At large $g$, the two lines on the left panel  are separated by a 
finite distance $\pi/2$ due a mismatch of terms proportional to $\pi g$ in \re{WeakOcta} and \re{InfinitGoctagon}. }
\label{fig:null}
\end{figure}

We recall that at strong coupling the relations \re{WeakOcta} and \re{InfinitGoctagon} differ by the terms proportional to $g\pi$ and they exhibit different dependence on $\xi$. We will 
verify them in two steps. First, we put  $\xi=0$ and examine the dependence 
of $\partial_g\log\mathbb O(0)$ on the coupling constant at $y=10$. Second, in order to check the $\xi-$dependence, we consider the logarithmic derivative of the ratio of octagons 
$\partial_g\log(\mathbb O(0)/\mathbb O(\xi)) $  
and examine its $g-$dependence for $\xi=1/10$. We observe   (see Figure~\ref{fig:null}) that the relations \re{WeakOcta} and \re{O-impr} are in agreement with numerical values at 
lower and higher values of $g$, respectively. The transition between the two regimes occurs for small $g'=g-L/(2\pi)$, or equivalently for $g\sim \log y/(2\pi)$.

\section{Conclusions}
\label{SectConclusion}

In this paper, we demonstrated that the correlation function of four infinitely heavy half-BPS operators defined in \re{G4} satisfies a system of nonlinear integro-differential equations in planar 
$\mathcal N=4$ SYM. These equations are powerful enough to determine the correlation function  for arbitrary values of the 't Hooft coupling and for generic values of the cross ratios. 

The starting point of our consideration was a representation of the correlation function as a determinant of a semi-infinite matrix previously derived in Refs.~\cite{Kostov:2019stn,Kostov:2019auq} 
using integrability-based hexagonalization framework~\cite{Basso:2015zoa,Fleury:2016ykk,Eden:2016xvg}. The matrix in question describes magnons propagating on a worldsheet of the octagon 
in the dual string theory and it has interesting properties. We found that it can be brought by an appropriate similarity transformation to a block-diagonal form. This allowed us to express the correlation 
function as a Fredholm determinant of an integral operator on a half-line with a well-known Bessel kernel modified by a cut-off Dirac-Fermi-like function. The resulting Fredholm determinant 
representation of the four-point correlation function has a striking similarity to two-point functions in integrable low-dimensional integrable models. Taking advantage of this fact, we applied the 
method of differential equations developed in Ref.~\cite{Its:1990} and derived a system of nonlinear equations for the octagon. 

At weak coupling, a solution to these equations yields the known expansion of the octagon over ladder integrals. At strong coupling, we developed a systematic expansion of the octagon in the inverse 
powers of the coupling constant and derived a representation of the corresponding coefficients as integrals of the cutoff function. We examined the resulting strong coupling expansion of the correlation 
function in various kinematical regions and observed a perfect agreement both with its expected asymptotic behavior dictated by the OPE as well as outcomes of numerical evaluation of the octagon. We 
found that, surprisingly enough, the strong coupling expansion is Borel summable. Applying the Borel-Pad\'e method, we demonstrated that the improved strong coupling expansion correctly describes 
the correlation function over a wide region of the coupling constant. 

There are several avenues for further studies. 

Analyzing the strong coupling expansion of the octagon, we focused on perturbative corrections in $1/g$ and systematically discarded exponentially small nonperturbative corrections suppressed by 
powers of $\e^{- \pi g}$. The latter are ubiquitous to strong coupling analyses in AdS/CFT and it would be interesting to elucidate their origin. Recall that in the well-studied case of the cusp anomalous 
dimension, nonperturbative effects manifest themselves through Borel singularities of the strong coupling expansion in $1/g$.  They are associated with a formation of a mass gap in the nonlinear O(6) sigma 
model describing massless excitation of the string worldsheet~\cite{Alday:2007mf,Basso:2009gh}. Particularly, it would  be important to reconcile exponentially small corrections to the octagon with the 
apparent fact of Borel summability of the corresponding perturbative series in $1/g$. 

According to AdS/CFT, the leading term $A_0$ in the strong coupling expansion of the octagon \re{O-str} should be given by a minimal area of a string worldsheet residing on four AdS geodesics. It would 
be very interesting to compute $A_0$ directly from string theory. Quantum fluctuations of the string generate corrections to \re{O-str} suppressed by powers of $1/g$. It would be even more exciting to 
reproduce the relations \re{u's} and, in particular, unravel the origin of universality of $\log g$ enhanced term in \re{O-str} coming from quadratic fluctuations. The method of differential equations allowed 
us to compute all the coefficients in \re{O-str} except the constant, $g-$independent term $B$ which appears as an integration constant in \re{D-u}.  One can use the relation \re{d-const} to find its 
dependence on the kinematical variables but the calculation turns out to be surprisingly complicated. The main reason is that the integration in \re{d-const} does not commute with strong coupling expansion 
of the function $Q(x)$. Namely, all terms in $1/g$ expansion of this function contribute to $B$ and the strong coupling series needs to be resummed to all orders in $1/g$. This question deserves a thorough 
investigation.

Intriguingly, the anomalous dimension $\Gamma (g)$ describing the leading asymptotic behaviour of the null octagon \re{WeakOcta}
also controls a double-logarithmic behavior of the six-gluon MHV amplitude in a  kinematical 
limit when three adjacent pairs of gluon momenta become collinear simultaneously \cite{Basso:2020xts}. 
We also observed in Ref.~\cite{Belitsky:2019fan} that this anomalous dimension admits a complimentary description in terms of a flux-tube-like equation, whose origin remains obscure to us. 
These two facts hint towards the existence of a nontrivial connection between tessellation of amplitudes in terms of pentagons, on the one side, and hexagonalization of correlators, on the other.

It would be interesting to extend the formalism developed in this paper to more complicated correlation functions. The most immediate example are the octagons with nonzero internal 
bridges~\cite{Coronado:2018ypq}. They admit a Fredholm determinant representation similar to the one studied above~\cite{Kostov:2019stn,Kostov:2019auq}
and determine the so-called asymptotic correlation functions  \cite{Basso:2017khq,Coronado:2018ypq}. We plan to address these questions in the future.

\section*{Acknowledgments}

We would like to thank Ivan Kostov, Paul Krapivsky, Valentina Petkova, Alexander Povolotsky and Didina Serban for interesting discussions, and Till Bargheer and  Pedro Vieira for careful reading of 
the manuscript. We are grateful to Riccardo Guida for his generous help with numerical calculations.
The research of A.B.\ and G.K.\ was supported, respectively, by the U.S. National Science Foundation under the grant PHY-1713125 and by the French National Agency for Research grant ANR-17-CE31-0001-01.

\appendix
 
\section{Analytical regularization}\label{app:reg}

In this appendix, we discuss the strong coupling expansion of the following integral
\begin{align}\label{Jn}
f_n(g) 
=\int_0^\infty dz  \,  z^{2n}  \partial_{z}\log(1-\widehat \chi (z)) J_1^2(2gz)\,,
\end{align}
where the cut-off function $\widehat \chi (z)$ is given by \re{chi}. It arises as the leading term in the expansion of the moment $Q_n=-\int_0^\infty dz\,z^{2n} q(z) \partial_z \widehat \chi(z)$ after we 
replace the function $q(z)$ with its expression \re{q-gen} to the leading order in $1/g$.

We perform the calculation of \re{Jn}  using two different methods. The first one is based on the well-known Mellin-Barnes representation of the Bessel functions. The second one employs an analytical regularization 
described in Sect.~\ref{sect:reg}. We show that the two methods give the same result.

To start with, recall the Mellin-Barnes transform of the product of Bessel functions in \re{Jn} 
\begin{align}
J_1^2(2gz) = \int {dj\over 2\pi i} (2gz)^{-j} \varphi(j)\,,\qquad 
\varphi(j) = \frac{\Gamma \left(\frac{1}{2}-\frac{j}{2}\right) \Gamma \left(\frac{j}{2}+1\right)}{2
   \sqrt{\pi } \Gamma \left(1-\frac{j}{2}\right) \Gamma \left(2-\frac{j}{2}\right)}
   \, ,
\end{align}
where the integration contour runs parallel to the imaginary axis and $-2 <  {\rm Re}\, j < 1$. Taking into account this relation, we get from \re{Jn}
\begin{align}\label{Jn-poles}
f_n(g) 
=  \int {dj\over 2\pi i} (2g)^{-j} \varphi(j) \int_0^\infty dz  \,  z^{2n-j}  \partial_{z}\log(1-\widehat \chi (z))\,.
\end{align}
At large $g$, we can deform the integration contour to the right towards large positive $j$ and pick up residues at poles encountered along the way. The poles emerge  from the function $\varphi(j)$ as well as 
from the $z-$integral. Similar integrals had already occurred in Eq.\ \re{I-anal}. We recall that $ \partial_{z}\log(1-\widehat \chi (z))= 2/z +O(z)$ for  $z \to 0$ and, therefore, the integration over small $z$ produces 
poles at $2n-j=0,-2,-4,\dots$, or equivalently $j=2n,2(n+1),2(n+2),\dots$.~\footnote{Here we took into account that $\widehat\chi(z)$ is an even function of $z$.} Notice however that $\varphi(j)$ vanishes for 
$j=2,4,\dots$ and, therefore, the integral \re{Jn-poles} has zero residues at all poles mentioned above except $j=0$. Such pole only appears for $n=0$. 

In this case, we have
\begin{align}\notag
f_0(g) = & - \varphi(0) \res_{j=0} \int_0^\infty dz  \,  z^{-j}  \partial_{z}\log(1-\widehat \chi (z))
\\
& -\sum_{k\ge 1} (2g)^{-2k+1} \res_{j=2k-1} \varphi(j) \left[ \int_0^\infty {dz\over z^{2k}}  \,  z  \partial_{z}\log(1-\widehat \chi (z))\right]_{\rm reg}\,.
\end{align}
Here in the second line, the sum runs over the poles of the function $\varphi(j)$ and $[\dots]_{\rm reg}$ denotes the regular part of the integral. Similarly to \re{I-anal} and \re{profile1}, it can be found by 
integrating by parts $k$ times. In this manner, we obtain
\begin{align}\label{f0}
f_0(g) = 1 - \sum_{k\ge 1}(-1)^k { I_k \over (2g)^{2k-1}} \frac{\sqrt{\pi }  \Gamma \left(k+\frac{1}{2}\right)}{\Gamma
   \left(\frac{3}{2}-k\right) \Gamma \left(\frac{5}{2}-k\right) \Gamma (k)}\,,
\end{align}
where $I_k$ is the profile function defined in \re{profile1}.

For $n\ge 1$, the integral in \re{Jn-poles} is given by the sum over residues of the function $\varphi(j)$ 
\begin{align}\notag\label{fn-exp}
f_n(g)   
&= -\sum_{k\ge 1} (2g)^{-2k+1} \res_{j=2k-1} \varphi(j) \left[ \int_0^\infty {dz\over z^{2k-2n}}  \,  z  \partial_{z}\log(1-\widehat \chi (z))\right]_{\rm reg}
\\
&=- \sum_{k\ge 1} {(-1)^k  \over (2g)^{2k-1}} \frac{\sqrt{\pi }  \Gamma \left(k+\frac{1}{2}\right)}{\Gamma
   \left(\frac{3}{2}-k\right) \Gamma \left(\frac{5}{2}-k\right) \Gamma (k)} I_{k-n} 
   \, ,
\end{align}
where $I_{k-n}$ is defined in \re{profile1}.

Let us now reproduce the same relations using an analytical regulation. Assuming that the dominant contribution to \re{Jn} comes from $z=O(g^0)$, we replace the Bessel function in \re{Jn} with 
its  asymptotic behavior at large $g$ 
\begin{align}
J_1(2gz)
   &={1\over\sqrt{2\pi g z}}\left[a_0(g z)\sin(2gz) + b_0(gz)\cos(2gz)\right]\,.
\end{align}
The explicit expressions for $a_0$ and $b_0$ can be read from \re{hat-q-as}. Replacing rapidly oscillating functions in the expression for $J_1^2(2gz)$ with their mean values, we get from  \re{Jn}
\begin{align}\label{psi}
f_n(g)   
= {1
\over 2\pi g} \int_0^\infty dz  \,  z^{2n-1}  \partial_{z}\log(1-\widehat \chi (z))\times \frac12  \left[a_0^2(g z) + b_0^2(gz) \right] \,,
\end{align}
where
\begin{align}
\frac12  \left[a_0^2(g z) + b_0^2(gz) \right]
=1 +\frac{3}{32 g^2 z^2} -\frac{45}{2048 g^4 z^4} +
\frac{1575}{65536 g^6 z^6} + O(1/(gz)^8)\,.
\end{align}
Expanding \re{psi} in powers of $1/g$, we encounter the same integral as in \re{profile}. We use the analytical regularization to 
define it according to \re{profile1}. This leads to
\begin{align}\label{fn-id}
f_n(g)   
= {1
\over 2 g}\left[ I_{1-n} +\frac{3}{32 g^2}I_{2-n} -\frac{45}{2048 g^4}I_{3-n} +
\frac{1575}{65536 g^6}I_{4-n} + \dots \right]\,.
\end{align}
It is straightforward to verify that this relation coincides with \re{fn-exp} for $n\ge 1$.
However, for $n=0$ the relation \re{fn-id}  does not capture the leading, $O(g^0)$ term in the right-hand side of \re{f0}.   
The missing contribution comes from the integration over the region $z=O(1/g)$. Indeed, restricting the integration in \re{Jn} to $0\le z<c/g$, we get
\begin{align} 
f_0 & =\int_0^{c/g} dz  \, \partial_{z}\log(1-\widehat \chi (z)) J_1^2(2gz)+\dots
 =2\int_0^{ c} {dx \over x} J_1^2(2x)+ \dots  = 1+ \dots\,,
\end{align}
where the dots denote contributions from $z=O(g^0)$. Here in the second relation, we changed the integration variable to $x=gz$, took the limit $g\to\infty$ together with  $c\gg 1$. 

The above analysis demonstrates that the analytical regularization captures the contribution to \re{Jn} coming from integration over $z=O(g^0)$ and, therefore, 
suppressed by powers of $1/g$. The remaining $O(g^0)$ contribution comes from $z=O(1/g)$ and it is controlled by the behavior of the function $q(z)$ at small $z$.

\section{Quantization condition from zeroth moment }\label{app:sol}

We demonstrated in Section~\ref{sect:strong} that the relation \re{u-chain}, combined with the strong coupling expansion of the function \re{q-gen} for $z=O(g^0)$,  
can be used to compute the octagon \re{O-str} at strong coupling. 

In this appendix, we present another approach to computing the octagon. It relies on the relation \re{Q0-mom} and makes use of the strong coupling expansion of the function \re{q-gen} 
for $z=O(1/g)$. As we demonstrate below, it yields the same result for the expansion coefficients \re{u's} and allows us to establish the relations \re{u-odd} between the coefficients with 
odd and even indices.  

We introduce $x=2gz$ and look for the solution to \re{diffeq} in the form \re{q-gen} with
\begin{align}\label{f-exp}
f= f_0(x) + {1\over g} f_1(x) +{1\over g^2} f_2(x) +\dots\,,
\end{align}
where $x=O(g^0)$ and $g\gg 1$. Substituting this ansatz into \re{diffeq} and matching the coefficients in front of powers of $1/g$ we obtain the system of differential equations
\begin{align}\notag\label{system}
& x^2 f_0''(x)+x f_0'(x)+( x^2-1)f_0(x) =0 \,,
\\[3mm]\notag
& x^2 f_1''(x)-x   f_1'(x) + x^2 f_1(x)+A_2 f_0(x)=0 \,,
\\[3mm] 
&  x^2 f_2''(x)-3 x f_2'(x)+( x^2 +3)f_2(x)+ A_3 f_0(x)+A_2 f_1(x)=0 \,.
\end{align}
We verify that, in accord with \re{q-gen}, the solution to the first equation is 
\begin{align}\label{f0-J}
f_0(x) = J_1(x) \,.
\end{align}
Solving \re{system} we assume that the expansion \re{f-exp} is uniform at small $x$, i.e. $f_1$, $f_2,\dots$ do not modify the leading small $x$ behavior and scale as $f_n\sim x$ for $x\to 0$. 

At small $x$, we substitute $f_0(x) \sim x/2 $ into the second relation in \re{system} to find that $f_1(x) \sim A_2\, x/2$.
In a similar manner, replacing $f_2(x)\sim  c\, x$ in the last relation of \re{system}, we notice that the terms proportional to $c$ drop out resulting to
\begin{align}
A_3 f_0(x) + A_2 f_1(x) = O(x^2) \,.
\end{align}
Equating to zero the coefficient in front of $x$ on the left-hand side we arrive at $A_3+A_2^2=0$, in agreement with \re{u-odd}.

The same mechanism is at work for high-order  functions in \re{f-exp}. Replacing $f_\ell(x)= \sum_{k\ge 1} f_{\ell,k} x^k$ in \re{f-exp} we can obtain from 
\re{diffeq} the system of linear equations for the coefficients $f_{\ell,k}$. For $\ell=4,6,\dots$ this system turn out to be overdetermined. For the solution to exist the coefficients with odd indices $A_{2k+1}$ 
have to satisfy consistency conditions. They take the form \re{u-odd} and allow us to express $A_{2k+1}$ in terms of the coefficients with even indices. To determine even coefficients $A_2, A_4,\dots$, 
we use the relation for zeroth moment \re{Q0-mom}.

Substituting \re{q-gen} and \re{f-exp} into \re{Q0-mom}, we obtain the relation
\begin{align}\label{J-eq}
Q_{0,0} +{1\over g} Q_{0,1} + {1\over g^2} Q_{0,2} = O(1/g^3)\,,
\end{align}
where the notation was introduced for 
\begin{align}\notag\label{QQ}
& Q_{0,0}=\int_0^\infty dz  \,    \partial_{z}\log(1-\widehat \chi (z)) f_0^2(gz) -1\,,
\\\notag
& Q_{0,1}=2\int_0^\infty dz  \,    \partial_{z}\log(1-\widehat \chi (z)) f_0(gz)f_1(gz)\,,
\\
& Q_{0,2}=\int_0^\infty dz  \,    \partial_{z}\log(1-\widehat \chi (z)) \lr{f_1^2(gz) + 2f_0(gz) f_2(gz)} \,.
\end{align}
We show below that these functions are given by series in $1/g$ with the coefficients depending on the expansion coefficients $A_k$ of the octagon \re{O-str}.  Equating to zero the coefficients in front of 
powers of $1/g$ on the left-hand side of \re{J-eq}, we can determine $A_k$.

The leading function $Q_{0,0}$ can be computed using the relations \re{Jn} and \re{fn-id}
\begin{align}\label{Q00}
Q_{0,0}= f_0-1=\frac{I_1}{2 g}  +O(1/g^3)\,,
\end{align}
where $I_1$ is given by \re{profile1}. According to \re{J-eq}, the $O(1/g)$ contribution to $Q_{0,0}$ should cancel against $O(g^0)$ contribution from $Q_{0,1}$ on the left-hand side of \re{J-eq}.

To compute  $Q_{0,1}$ we need an expression for the functions $f_1(x)$. Solving the second equation in \re{system} 
we get
\begin{align}
\label{f1}
f_1 (x) =    -\frac{\pi A_2}2 \left[ x J_1(x)  \int_x^\infty \frac{d{x'}}{ x'{}^2} Y_1({x'}) f_0({x'})
+x Y_1(x) \int_0^x\frac{d{x'}}{{x'}^2}J_1({x'})f_0({x'}) \right] ,
\end{align}
where $J_1$ and $Y_1$ are Bessel functions. Here the integration contours are chosen in such a way that the two integrals are convergent.
In general, $f_1$ is defined up to a solution to the homogenous equation, $f_1\to f_1+ c_1 x J_1(x) + c_1' x Y_1(x)$. The value of $c_1'=0$ is fixed from the requirement that $f_1=O(x)$ at small $x$. To fix $c_1$, we require that the contribution of $J_1(x)$ to $f_1(x)$ should not modify asymptotic behavior of the leading function \re{f0-J} at large $x$. This leads to $c_1=0$.

The leading, $O(g^0)$ contribution to $Q_{0,1}$ comes from the integration over small $z=O(1/g)$ in \re{QQ}.
Replacing ${\partial_z \log(1-\widehat \chi(z))} = 2/z + O(z)$ in \re{QQ} and changing the integration variable to $z=x/(2g)$ we evaluate the leading contribution to $Q_{0,1}$ as
\begin{align}\notag\label{Q01}
Q_{0,1} & ={4 }\int_0^\infty {dx\over x} J_1(x) f_1(x) +O(1/g^2)
 \\\notag
& = - 2\pi A_2 \int_0^\infty {dx}\, J_1(x) Y_1(x) \int_0^x {d{x'}}\, J^2_1({x'}) \lr{{1\over {x'}^2}+{1\over x^2}}+O(1/g^2)
\\
& = {2\over 3} A_2+O(1/g^2)\,.
\end{align}
Substituting \re{Q00} and \re{Q01} into \re{J-eq} we find that the coefficient in front of $1/g$ vanishes provided that
$A_2=-3 I_1/4$, in agreement with \re{u's}.

Moving on to the next order, the solution to \re{system} for $f_2$ is
\begin{align}\label{f2-sol}
f_2 (x) = - \frac{\pi}{2} z^2
\bigg[
&
J_1 (x) \int_x^\infty \frac{d {x'}}{{x'}^3} Y_1 ({x'}) [ A_3 f_0 ({x'}) + A_2 f_1 ({x'})]
\nonumber\\
-
&
Y_1 (x) \int_x^\infty \frac{d {x'}}{{x'}^3} J_1 ({x'}) [ A_3 f_0 ({x'}) + A_2 f_1 ({x'})]
+
c_2 J_1 (x) + c_2^\prime Y_1 (x)
\bigg]
\, .
\end{align} 
Again, we have two integration constants $c_2$ and $c_2^\prime$ accompanying the solutions to the homogenous equation. Taking into account that $f_0({x'})\sim {x'}/2$ and $f_1({x'})\sim A_2{x'}/2$ for small ${x'}$, 
it is easy to check that the integral on the second line of the last relation modifies the asymptotic behavior of $f_2(x)$ at small $x$, e.g. $f_2(x) \sim   (A_3+A_2^2)\,x \log x/4$. Requiring such terms do not 
appear, we find that $A_3=-A_2^2$, independently of the values of $c_2$ and $c_2'$. 

The coefficients $c_2$ and $c_2'$ can be determined from relation \re{J-eq}. Notice that in virtue of \re{Q00} and \re{Q01}, the first two terms in the left-hand side of \re{J-eq} do not produce $O(1/g^2)$ 
corrections and, therefore, the relation \re{J-eq} leads to $Q_{0,2}=0+O(1/g)$. Substituting \re{f2-sol} into \re{QQ} and going through a rather elaborate calculation, we find that $c_2^\prime Y_1 (x)$ term 
in \re{f2-sol} produces the contribution to $Q_{0,2}$ that scales as $O(g)$. To avoid it we require that $c_2'=0$. Then, the vanishing of $O(g^0)$ correction to $Q_{0,2}$ leads to 
$c_2 = 16 A_2^2/9$.  
 
The above analysis can be extended to high orders in $1/g$. It leads to the same result \re{u's}, but calculations become more involved and tedious as compared to those presented in Section~\ref{sect:strong}. 
 
\section{Matrix representation}\label{app:mat}
 
In this appendix, we present a solution to the differential equation \re{sys} in terms of a semi-infinite matrix \re{oct1}.

According to \re{Q-def}, the function $Q(x)$ admits a representation
\begin{align}\label{Q-def1} 
 Q(x)  = 
 \phi(x) + \vev{x| {\mathbb{K}_\chi \over 1- \mathbb{K}_\chi} |\phi} 
 = \phi(x) + \sum_{\ell\ge 1}\vev{x| (\mathbb{K}_\chi )^\ell |\phi} \,,
\end{align} 
where $\phi=J_0(\sqrt{x})$ and the operator $\mathbb{K}_\chi$ is defined in \re{calK}. To evaluate the second term in \re{Q-def1}, we first determine the kernel of the integral operator $(\mathbb{K}_\chi)^\ell$
\begin{align}
\vev{x|  (\mathbb{K}_\chi)^\ell |x'} = \int_0^\infty dx_1 \dots d x_{n-1} K(x,x_1) \chi(x_1) \dots K(x_{n-1},x') \chi(x')\,.
\end{align}
Replacing the $K-$kernel with its explicit expression (see the first relation in \re{K-JJ}), we get 
\begin{align}
\vev{x| (\mathbb{K}_\chi)^\ell |x'} &= \sum_{n,m =0}^\infty (2m+1) {J_{2n+1}(\sqrt{x})\over \sqrt{x}} (-1)^{n+m}(k_-^{\ell-1})_{nm}
 {J_{2m+1}(\sqrt{x'})\over \sqrt{x'}} \chi(x') \,,
\end{align}
where the matrix $k_-$ is given by \re{oct1}. Substitution of this relation into \re{Q-def1} yields the following representation for the solution to \re{sys}
 \begin{align} \label{Q-rep}
Q(x) = J_0(\sqrt{x}) + \sum_{n,m =0}^\infty  {J_{2n+1}(\sqrt{x})\over \sqrt{x}} (-1)^{n}\lr{1\over 1- k_-}_{nm}
(2m+1)\,  \chi_m\,.
\end{align}
Here the notation was introduced for 
\begin{align}\notag\label{chi-rep}
 \chi_m 
&= (-1)^m  \int_0^\infty {dx'\over \sqrt{x'}} \chi(x') J_{2m+1}(\sqrt{x'}) J_0(\sqrt{x'})
\\
&= 4g (-1)^m  \int_0^\infty {dz}{J_{2m+1}(2gz) J_0(2gz) }\widehat\chi(z)\,,
\end{align}
where $x'=2g z$ and the function $\widehat\chi(z)$ is defined in \re{chi}. 
 
For the potential we find in a similar manner from \re{u-def}
\begin{align}\notag\label{u-rep}
u & 
=\vev{\phi|\chi  |\phi}  + \vev{\phi|\chi {K \chi\over 1- K \chi} |\phi} 
\\
& = \int_0^\infty dx \, \chi(x) J_0^2(\sqrt{x}) +  \sum_{n,m =0}^\infty   \chi_n \lr{1\over 1- k_-}_{nm}
(2m+1)  \chi_m\,.
\end{align}
It follows from the second relation in \re{chi-rep} that $\chi_m=O(g^{2m+2})$ at weak coupling.  As a consequence, the weak coupling expansion of \re{Q-rep} and \re{u-rep} 
 contains a finite number of terms at any order in the coupling. 
 
 At finite coupling, we can approximate the exact expressions for $Q(x)$ and $u$ by retaining  a sufficiently large number of terms  in \re{Q-rep} and \re{u-rep}. To this end, we 
 replace the semi-infinite matrix $(k_-)_{nm}$  in \re{Q-rep} and \re{u-rep} by its finite-dimensional minor, $n,m\le N_{\rm max} \sim 10^2$. This procedure proves to be efficient in numerical analysis of the 
 differential equation \re{sys}.
 
 \section{Profile function in the null limit}\label{app:prof}
 
Expanding the cut-off function \re{chi} at small $\xi$ and large $y$, we get
\begin{align}\notag
z\partial_z \log(1-\widehat\chi) &=  {z\over 1+\e^{z-y}} + {2z\over  \e^{z}-1}
+\xi^2 \bigg[ \frac{1}{2 \left(1+\e^{z-y}\right)^2} -\frac{z+1}{2 z
   \left(1+\e^{z-y}\right)}
\\   
  &  +\frac{2
   z}{\left(\e^z-1\right)^3}+\frac{3 z-1}{\left(\e^z-1\right)^2}+\frac{z^2-z-1}{\left(\e^z-1\right) z}
   \bigg] +O(\xi^4)+ O(\e^{-y})\,.  
\end{align}
Substituting this relation into \re{profile1}, we obtain for the profile function
\begin{align}\label{I-exp1}
I_n=I_n^{(0)} + \xi^2 I_n^{(1)} + \xi^4 I_n^{(2)} + O(\xi^6)\,.
\end{align}
For the leading function, the calculation gives
\begin{align}\notag
& I_0^{(0)} ={1\over2\pi}\left(\pi^2 + y^2 \right)\,,\qquad
\\ \notag & 
I_1^{(0)} =  {1\over \pi} \lr{\log y +\gamma-\log(2\pi)} \,,
\\ & 
I_n^{(0)} =(-1)^{n+1}   {2\zeta (2 n-1)\over (2\pi)^{2n-1}}\,, \qquad \text{for $n\ge 2$}\,,
\end{align}
where $\gamma$ is the Euler-Mascheroni constant. Here the first relation is exact whereas the remaining ones hold up to corrections suppressed by powers of $1/y^2$. Note that the 
functions $I_1^{(0)}$ and $I_n^{(0)}$ possess weight $0$.

For the $O(\xi^2)$ and $O(\xi^4)$ corrections to the profile function, we find in the same fashion
\begin{align}\notag
& I_0^{(1)} = -{1\over 2\pi}\lr{\log y +1 +\gamma-\log(2\pi)}\,,
\\ \notag
& I_{k+1}^{(1)} =
(-1)^{k+1}  (2 k+1)  (2 k+3)  {\zeta (2 k+3)\over  (2\pi)^{2k+3 }}\,, 
\\
& I_k^{(2)} =\frac{1}{12} (-1)^{k+1} (2 k-1) (2 k+1) (2 k+3) (2 k+5)  {\zeta (2k+3)\over (2\pi)^{2k+3}}\,, 
\end{align}
where $k\ge 0$. Higher order corrections to the profile function \re{I-exp1} are given by 
\begin{align}
I_k^{(p)} =  (-1)^{k+p-1} {2\zeta (2k+2p-1)\over (2\pi)^{2k+2p-1}} {(2k+4p -3 )!!\over (2k-3)!!  (2p)!}\,, \qquad \text{for $k\ge 0$ and $p\ge 2$}\,.
\end{align} 
The above relations are valid up to corrections vanishing as $y\to\infty$. 

According to  \re{I-exp1}, the profile function $I_n$ depends on $y$ only for $n=0$ and $n=1$. Moreover, this dependence is confined to the first two terms in the $\xi^2-$expansion.

\bibliographystyle{JHEP} 

\bibliography{papers}

\end{document}